\documentclass[sigconf]{acmart}
\usepackage{amsthm, bm}
\usepackage{blkarray}
\usepackage{amsmath}
\usepackage{sidecap}
\usepackage{enumitem}
\usepackage{subfigure}
\usepackage{subcaption}
\usepackage{pifont}
\usepackage{hyperref}

\usepackage{algpseudocode}
\usepackage[ruled,vlined,linesnumbered]{algorithm2e}
\usepackage{cancel}
\def\cred{\textcolor{red}}

\algrenewcommand\algorithmiccomment[1]{\textcolor{olive}{\footnotesize\textit{// #1}}}

\newcommand{\new}[1]{#1}

\AtBeginDocument{%
  \providecommand\BibTeX{{%
    \normalfont B\kern-0.5em{\scshape i\kern-0.25em b}\kern-0.8em\TeX}}}

\setcopyright{acmlicensed}
\acmJournal{PACMMOD}
\acmYear{2025} \acmVolume{3} \acmNumber{3 (SIGMOD)} \acmArticle{183} \acmMonth{6}\acmDOI{10.1145/3725412}
\acmConference[SIGMOD '25]{ACM SIGMOD Conference on Management of Data}{June 22--27,
   2025}{Berlin, Germany}

\begin{document}

\title{Low Rank Learning for Offline Query Optimization}
% Look Ma, No Features! Steered Query Optimization with Linear Methods
% No features, no network, no problem! Steered query optimization with Linear Methods
% Mo' weights, mo' problems: steered query optimization with linear methods

\author{Zixuan Yi}
\orcid{0009-0001-8015-8486}
\affiliation{%
\institution{University of Pennsylvania}
\country{USA}
}
\email{zixy@cis.upenn.edu}

\author{Yao Tian}
\orcid{0000-0001-6876-5059}
\affiliation{%
  \institution{The Hong Kong University of Science and Technology}
  \country{China}
}
\email{ytianbc@cse.ust.hk}

\author{Zachary G. Ives}
\orcid{0000-0001-7527-2957}
\affiliation{%
  \institution{University of Pennsylvania}
  \country{USA}
}
\email{zives@cis.upenn.edu}

\author{Ryan Marcus}
\orcid{0000-0002-1279-1124}
\affiliation{%
  \institution{University of Pennsylvania}
  \country{USA}
}
\email{rcmarcus@cis.upenn.edu}
%\author{Anonymous authors for review}

\renewcommand{\shortauthors}{Yi et al.}
%\renewcommand{\shortauthors}{Anonymous authors for review}

\begin{abstract}

% Recent advances in learned query optimization have shown significant performance improvements, but also come with challenges such as unpredictable regressions, high training costs, and dependencies on specific DBMS features. To address these issues in repetitive analytic workloads, we propose \lime, a framework that uses offline exploration to minimize resource usage and maintain performance stability. 

Recent deployments of learned query optimizers use expensive neural networks and ad-hoc search policies. To address these issues, we introduce \lime, a framework for offline query optimization leveraging low-rank learning to efficiently explore alternative query plans with minimal resource usage.
By modeling the workload as a partially observed, low-rank matrix, we predict unobserved query plan latencies using purely linear methods, significantly reducing computational overhead compared to neural networks. 
% Unlike previous approaches that place expensive neural networks directly in the query processing ``hot'' path, \lime offers a simpler, more efficient solution without assumptions about the underlying DBMS. 
We formalize offline exploration as an active learning problem, and present simple heuristics that reduces a 3-hour workload to 1.5 hours after just 1.5 hours of exploration. 
%To handle long-running queries, we apply censored observations techniques.
Additionally, we propose a transductive Tree Convolutional Neural Network (TCNN) that, despite higher computational costs, achieves the same workload reduction with only 0.5 hours of exploration.
Unlike previous approaches that place expensive neural networks directly in the query processing ``hot'' path, our approach offers a low-overhead solution and a no-regressions guarantee, all without making assumptions about the underlying DBMS. 
The code is available in \href{https://github.com/zixy17/LimeQO}{https://github.com/zixy17/LimeQO}.
%our approach avoids dependencies on specific database management system features and maintains a no-regressions guarantee, offering a generalizable and low-overhead solution for offline query optimization.
% By modeling offline optimization as a matrix completion problem, \lime predicts query performance with 100x fewer computational resources. Our experiments show that \lime can accelerate a 3-hour workload by a factor of 2 with only a few hours of offline exploration, making it a compelling addition to learned optimization techniques.

% We present \lime, a learned steering query optimizer based on \underline{li}near \underline{me}thods, such as matrix completion, for repetitive workloads. \lime can forgo expensive neural networks by taking advantage of the low-rank structure of query workloads. Using offline execution, \lime can accelerate workloads by up to 2x with zero regressions in just a few hours, while using 100-1000x fewer computational resources than deep learning techniques.
\end{abstract}

%\begin{CCSXML}
%\end{CCSXML}

%\ccsdesc[500]{Do Not Use This Code~Generate the Correct Terms for Your Paper}
%\ccsdesc[300]{Do Not Use This Code~Generate the Correct Terms for Your Paper}
%\ccsdesc{Do Not Use This Code~Generate the Correct Terms for Your Paper}
%\ccsdesc[100]{Do Not Use This Code~Generate the Correct Terms for Your Paper}


%\keywords{Do, Not, Us, This, Code, Put, the, Correct, Terms, for, Your, Paper}


\received{October 2024}
\received[revised]{January 2025}
\received[accepted]{February 2025}

\newcommand{\sparagraph}[1]{\vspace{2mm} \noindent \textbf{#1} } 
\newcommand{\wl}{ \mathbf{W}}
\providecommand{\W}{\mathbf{W}}
\providecommand{\A}{\mathbf{Q}}
\providecommand{\B}{\mathbf{H}}
\providecommand{\h}{h}
\providecommand{\qhp}{\{(q_{i_1},\h_{j_1}),\ldots,(q_{i_m},\h_{j_m})\}}
\providecommand{\R}{\mathbb{R}}
\providecommand{\WObserve}{\Tilde{\mathbf{W}}}
\providecommand{\WPredict}{\hat{\mathbf{W}}}
\providecommand{\WTimeout}{\Tilde{\mathbf{W_t}}}
\providecommand{\T}{\mathbf{T}}
\providecommand{\TExplore}{T_\text{Explore}}
\providecommand{\Mask}{\mathbf{M}}

\newcommand{\circleOne}{\ding{182}\xspace}
\newcommand{\circleTwo}{\ding{183}\xspace}
\newcommand{\circleThree}{\ding{184}\xspace}
\newcommand{\circleFour}{\ding{185}\xspace}
\newcommand{\circleFive}{\ding{186}\xspace}

\newcommand{\lime}[0]{\textsc{LimeQO}\xspace}
\newcommand{\limeplus}[0]{\textsc{LimeQO+}\xspace}
\newcommand{\greedy}[0]{\textsc{Greedy}\xspace}

\newcommand{\rand}[0]{\textsc{Random}\xspace}
\newcommand{\tcnn}[0]{\textsc{TCNN}\xspace}
\newcommand{\Bao}[0]{\textsc{Bao}\xspace}
\newcommand{\BaoCache}[0]{\textsc{Bao-Cache}\xspace}
\newcommand{\QOAdvisor}[0]{\textsc{QO-Advisor}\xspace}
\newcommand{\rcm}[1]{\textcolor{red}{[ Ryan: #1 ]}}
\newcommand{\zi}[1]{\textcolor{blue}{[ Zack: #1 ]}}

\def\cred{\textcolor{red}}
\def\cgreen{\textcolor{teal}}

\maketitle

\section{Introduction}

Recent advances in learned query optimization --- using machine learning to completely replace or aid a traditional query optimizer ~\cite{systemr} --- have demonstrated significant performance gains ~\cite{lero,balsa,fastgres,bao,hybrid_lqo}. However, learned optimizers also have several drawbacks: (1) the nature of learning techniques can cause \emph{unpredictable regressions} (e.g., ``my query was fast yesterday, why is it slow today?''), (2) they suffer from \emph{expensive training and inference costs}~\cite{low_data_lqo_benchmark} (e.g., from neural networks~\cite{neo} or from training data collection times~\cite{fastgres}), and (3) they often \emph{make assumptions about the underlying DBMS}, such as the availability of features (e.g., cost estimates~\cite{bao}) or the structure of query plans (e.g., tree structured plans with finite operators~\cite{balsa}).

In the context of repetitive analytic workloads, such as updating live dashboards and timely report generation, two recent works in production systems have addressed the first issue of unpredictable performance regressions: AutoSteer~\cite{autosteer} and QO-Advisor~\cite{bao_scope2}. The core idea behind both approaches is to use offline execution to verify that potential new query plans are actually better than the default plan. If verified, the new query plan is added to a plan cache and used when an eligible query arrives. This simple technique ensures that no query \emph{ever} regresses (absent data shift), but at the cost of potentially expensive offline execution.

%\zi{We should try to make this sound more impressive.  Maybe start with something like, "The broader problem of how to effectively explore the space of novel plans offline, while also avoiding regression, has not been systematically explored."}

\begin{SCfigure}
\centering
\begin{blockarray}{ccccc}

           & $h_1$    & $h_2$    & \dots    & $h_k$ \\
\begin{block}{c[cccc]}

  $q_1$    & 3        &     4    &   \dots  & ? \\
  $q_2$    & 9        &     ?    &   \dots  & 6 \\
  $q_3$    & ?        &     15   &   \dots  & 3 \\
  $\vdots$ & $\vdots$ & $\vdots$ & $\ddots$ & $\vdots$ \\
  $q_n$ & 5 & 1     & \dots & 2 \\
\end{block}
\end{blockarray}
\caption{An example \emph{workload matrix}. Each row represents a query, and each column represents a hint. 
%The entry in the $i$th row and $j$th column represents the performance of query $q_i$ with hint $h_j$. 
The value $?$ represents an unobserved latency. 
%Prior steering techniques implicitly try to complete this matrix.
}

\label{fig:ex-matrix}
\end{SCfigure}

While major steps in the right direction, neither AutoSteer nor QO-Advisor are deeply strategic in their offline exploration: both exhaustively test a set of alternative plans, and control for excessive offline exploration time by heuristically limiting the set of alternative plans explored. For example, QO-Advisor limits the set of alternative plans to those produced by a ``single rule flip''~\cite{bao_scope2}, while AutoSteer simply tests the $n$ most promising plans. To the best of our knowledge, \textbf{the broader problem of how to effectively explore the space of alternative plans offline, in a way that maximizes workload benefit while minimizing offline computation time, has not been systematically explored.}

\smallskip

Here, we formalize and expand on this offline exploration approach. Our proposed framework seeks to minimize offline resource usage while maximizing performance improvements, maintaining the ``no-regressions'' guarantee (compared to the underlying traditional optimizer) of prior work~\cite{autosteer,bao_scope2}. Additionally, we avoid any tight coupling between our framework and specific DBMSes: we do not make assumptions about the structure of query plans, the number of operators, or even the availability of cost estimates. The only assumption our framework makes is that each query plan has a number of alternatives with measurable latency. To accomplish this, we introduce a new approach to learned query optimization called \lime
%\footnote{The name of the system has been anonymized for review. We promise we thought of a sufficiently witty acronym.}
, which is trained with \emph{purely linear methods}, leading to drastically simpler and lower overhead implementations.

\sparagraph{Target workload \& hints.} Like AutoSteer and QO-Advisor, we target query workloads that are mostly repetitive.  Thus, we assume that \emph{most} queries and their set of potential query plans are known ahead of time, although we do support the addition of new queries over time. Additionally, like prior work on learned query optimization~\cite{bao, fastgres, autosteer, bao_scope, bao_scope2}, we assume that the underlying query optimizer provides a ``hint'' interface to create different variations of query plans. We justify these decisions in Section~\ref{sec:sys_model}.

\sparagraph{Workload matrix.}  Our core insight is to model the problem of offline optimization as a \emph{matrix completion (MC)}~\cite{als} problem: we can represent a workload with $n$ repeated queries and $k$ possible query hints as a \emph{workload matrix} $\W$, where each entry is the latency of a plan, as shown in Figure~\ref{fig:ex-matrix}. Selecting the best hint for each query amounts to taking a row-wise minimum. Unfortunately, computing the whole matrix would require $n \times k$ query executions, which could be prohibitive. Instead, we consider the workload matrix to be \emph{partially observed}: some entries are known (observed, or part of the training set), and other entries are unknown (unobserved). ``Completing'' the matrix means predicting the unobserved values.

Readers familiar with MC may note that one common application of MC is recommendation systems~\cite{mc_reco}. Indeed, we will show that \lime works for similar reasons as recommendation systems do: since sets of queries that perform well with some hints also tend to perform poorly with other hints, the \emph{rank} $r$ of the workload matrix is low. This \textbf{low rank} means that  (among other things) we can construct an accurate estimate of $\W$ using two factored matrices $\A \in \mathbb{R}^{n \times r}$ and $\B \in \mathbb{R}^{k \times r}$: $\hat{\W} = \A \B^T$. This approximation can be found using purely linear methods that use \emph{100x less computational resources than their neural network counterparts} (e.g., TCNNs~\cite{tree_conv}). 
%This reduces inference to the dot product of two $r$-dimensional vectors. %Intuitively, $\W$ is low rank because \emph{two queries that behave similarly on some hints are likely to behave similarly on other hints as well}; this means that a large portion of $\W$ can be explained by commonalities between queries and hints in the user's specific workload.

\sparagraph{Offline optimization as active learning.} Matrix completion allows us to approximate missing entries in the workload matrix, but we still need a way to explore the workload matrix efficiently. Ideally, we want to discover the minimum value of each row (the fastest hint for each query) as quickly as possible. This can be considered an \emph{active learning}~\cite{deep_active_learning_survey} problem, in which we must intelligently select which new pieces of information to observe next. Observing each new piece of information has an associated value (the amount we can improve the latency of the query) and a cost (the amount of offline exploration time we use). We present two simple algorithms inspired by active learning that are especially suited toward this special variant of the problem. These active learning techniques allow us to achieve near-optimal performance by spending only the default workload time (i.e., a few hours) rather than exhaustively exploring the entire space, which would take over 10 days.

\sparagraph{Transductive neural networks.} When computational overhead is not a limiting factor, and when certain assumptions (such as tree-structured plans) can be made about the underlying database system, our framework can also integrate expensive neural network models. Of course, doing so increases inference overhead, but may lead to faster convergence due to the power of neural networks. We present a new type of tree convolution neural network (TCNN)~\cite{tree_conv} called a \emph{transductive TCNN} which combines the tree-structured inductive bias of a TCNN with \emph{learned} representations of the $\A$ and $\B$ matrices. Unsurprisingly, our computationally expensive neural network is a better approximator of $\W$ than purely linear methods. It accelerates the 3-hour workload by a factor of 2 with 0.5 hours of offline exploration, whereas purely linear methods took 1.5 hours of exploration to achieve the same speedup. \new{However, the overhead of the neural network in the inference phase is 360 times higher than linear methods.}

\sparagraph{Trouble with timeouts.} A key challenge to exploring the workload matrix $\W$ is dealing with queries with unusually long latencies. For a particular row in the matrix, if the current best query plan takes $x$ seconds, then in some sense it is wasteful to execute any other plan in that row for longer than $x$ seconds: once a plan takes longer than $x$ seconds, we can rule it out as the optimal plan. Unfortunately, simply placing the timed out query value into $\W$ will mislead the machine learning model: it will look like the timed-out query plan took $x$ seconds to execute, but in reality, that plan could have taken much longer (the true latency of the plan is not known, but the fact that true latency is greater than $x$ is known). Prior work, like Balsa~\cite{balsa}, has addressed this issue by setting the query timeout to some integer multiple $S$ of $x$, allowing the model to at least see that the timed-out plan took longer than $Sx$ to execute. However, this solution is still suboptimal, since (1) it executes the timed-out query for longer than necessary (i.e., by an integer factor) and (2) still ``misleads'' the machine learning model by treating the latency of the query as $Sx$.

In this work, we show how a well-studied machine learning trick called \emph{censored observations}~\cite{survival_analysis} can be applied to learned query optimization. Using our technique, we can treat timed-out queries as ``first-class citizens,'' penalizing models for underestimating the the timed-out query's latency, but not penalizing the model for a (potentially valid) over-estimate. We show how to handle censored observations both in MC and in the transductive TCNN. Our experiment shows that the censored technique reduced the 3-hour workload to 1.5 hours after 0.5 hours of exploration, whereas without the censored technique, it took 0.9 hours to achieve the same reduction.

\sparagraph{Contributions} We make the following contributions:
\begin{itemize}[topsep=0pt,leftmargin = 9pt]
    \item We present \lime, a framework for offline exploration for query optimization formalized as an active learning problem, and present two simple heuristic solutions.
    \item We present a modified version of the popular alternating-least-squares (ALS)~\cite{als} MC algorithm which can handle censored observations (timeouts).
    \item We present the transductive TCNN, a neural network specially designed for offline query optimization which takes advantage of the low rank structure of the workload matrix, which can also handle censored observations.
    \item We show how \lime can be extended to handle new queries and data drift.
\end{itemize}

The rest of this paper is organized as follows. We present our system model in Section~\ref{sec:sys_model}. We define the offline exploration problem in Section~\ref{sec:offline_explore}. In Section~\ref{sec:expr}, we present experimental results. Finally, we present related works in Section~\ref{sec:related} and concluding remarks in Section~\ref{sec:conclusion}.

\section{Related Work} \label{sec:related}

\sparagraph{Learned Query Optimization.}
Recent works have explored the integration of machine learning techniques into several components in DBMS, such as learned cardinality estimators 
%\old{~\cite{deep_card_est2, robust_card_est, naru}} 
\new{~\cite{deep_card_est2, robust_card_est, naru, geo_ce, asm, gridar_lce, alece_lce}}, learned cost models ~\cite{zeroshot_latency_model}, and learned query optimizers.
Learned query optimizers are broadly divided into two categories: ``full'' learned optimizers that synthesize entire query execution plans from scratch, effectively replacing the traditional query optimizer ~\cite{rejoin,neo,balsa,roq,lero,sanjay_wat,qo_state_rep,loger,lstm_jo}, and ``steering'' learned optimizers that sit on top of a traditional optimizer~\cite{bao,fastgres,hybrid_lqo, lowrank,leon_qo}. The latter ``steering'' approach has fewer degrees of freedom, but exhibits lower variation, leading to adoption in some production systems~\cite{bao_scope2,autosteer,pilotscope}. Since any performance variation can be harmful to downstream applications, offline execution is often used to verify performance improvements~\cite{bao_scope2,autosteer,datafarm,hitthegym}, although confidence-learning based approaches are also being developed~\cite{eraser_lqo,roq,robopt}. Our approach builds upon the ``steering'' approach. \new{Unlike AutoSteer~\cite{autosteer} and QO-Advisor~\cite{bao_scope2}, we explore the space of query-hints combinations \emph{holistically}, taking the entire workload into account, reducing the need for exhaustive execution while still preventing regressions.} To the best of our knowledge, the only prior work on learned query optimization at the workload level is GALO~\cite{workload_reopt}, which mines query logs for problematic executions and recommends fixes for slow queries. Notably, GALO does not require offline query execution like the current work, but GALO is not guaranteed to be regression-free.

\new{Our approach is closely related to BayesQO~\cite{bayesqo}, a concurrent method also aimed at offline query optimization. However, while BayesQO optimizes one query at a time, our framework simultaneously optimizes an entire query workload. Furthermore, while BayesQO considers the entire set of possible query execution plans, our method considers a small, finite set of execution plans (e.g., 48 candidate plans in PostgreSQL). By optimizing at the workload level instead of the individual query level, our approach can leverage inter-query similarities and the low-rank structure of the workload matrix. Additional details on the performance differences are provided in Section ~\ref{sec:expr_bayesqo}.}

\sparagraph{Matrix Completion.} Matrix completion is a decades-old technique~\cite{goldberg1992using} that has been widely used in collaborative filtering~\cite{mf} and recommendation systems~\cite{als}, although MC has also been the subject of deep mathematical investigation~\cite{candes_power_2009}. Linear methods have also been used by learned cardinality estimators~\cite{leo}.  Several algorithms have been proposed for matrix completion, including nuclear norm minimization~\cite{nuclear}, singular value thresholding~\cite{svt}, and alternating least squares (ALS)~\cite{als}. More recently, deep learning techniques have been introduced to capture complex, non-linear relationships in recommendation systems~\cite{ncf, dnn_youtube_rec, dlrc_survey, dlrm, fm, deepfm}. These methods leverage neural networks to learn intricate patterns in user-item interactions, outperforming traditional linear models. Our \limeplus approach differs from these existing models by using the transductive approach ~\cite{transduction} that incorporate query plan trees into each matrix entry, rather than relying on the user and item features.

\sparagraph{Active Learning.} Active Learning  ~\cite{settles2009active,deep_active_learning_survey} is a well-established technique that aims to select the next training instance to label, with the primary goal of optimizing for the most information gain. As discussed in Section~\ref{sec:active-learning}, most existing works assume that the cost of acquiring the label is fixed~\cite{mc_queries, active_mc, activelearning_search}, irrespective of the instance, and thus does not need to balance between labeling cost and the value of the label. 
While Selective Supervision ~\cite{selective_supervision} does explore the idea of balancing the cost of labeling against  expected improvement, it still assumes a uniform cost within each class, which is not valid in our case.
% Moreover, at best we only have an estimate of the labeling cost and the value. 
Additionally, many multi-armed bandits-based active learning methods ~\cite{active_learning_pool, cmab_active_learning, active_learning_as_mab} assume an unlimited pool of observations, which also does not fit our scenario. 
We found these approaches ineffective for our matrix completion problem, highlighting the need for more customized strategies. 
Our \greedy strategy seeks to prioritize the longest-running query under the premise that it has the most potential value.  In contrast, \lime uses its current predictive model to estimate the instance with the largest expected benefit. 

% \zi{It might be useful to point to an active learning paper or two that we tried / considered, that didn't work?}
\section{System model} \label{sec:sys_model}

\begin{figure}[t]
    \centering
    \includegraphics[width=0.5\textwidth]{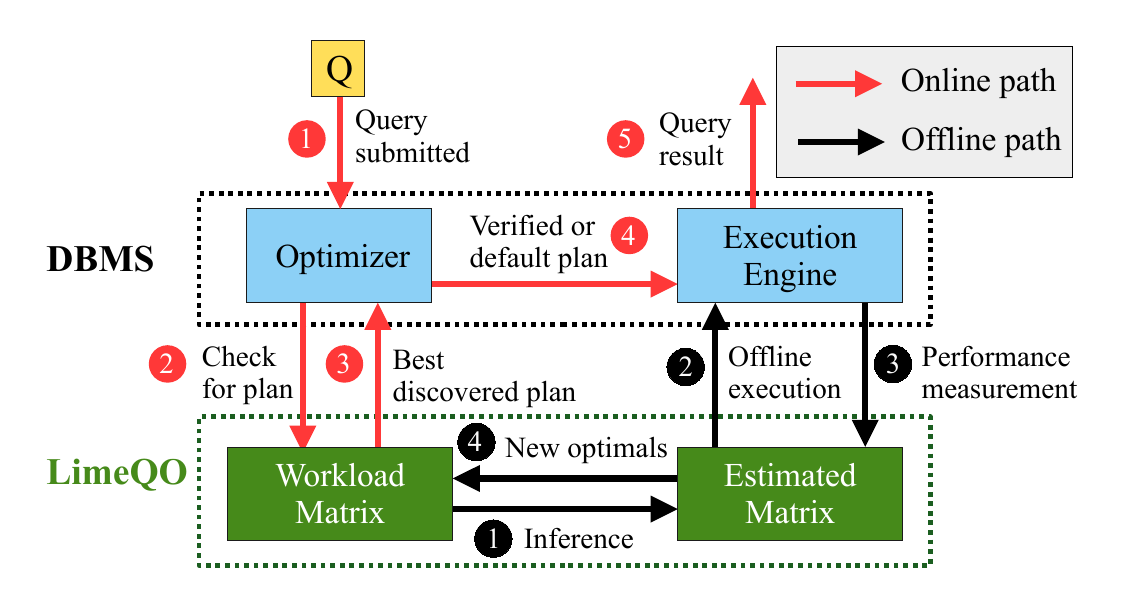}
    \caption{\lime system model}
    \label{fig:sysmod}
\end{figure}

\lime uses offline exploration to find the optimal hint for a set of queries in a repetitive workload. \lime does this by formulating the problem as low-rank matrix completion. Leveraging the low-rank property allows us to efficiently and accurately complete the workload matrix $\W$. 
\lime operates externally to the DBMS, interacting with both the query optimizer and the execution engine. It explores better plans within the available query hint space by ``steering''~\cite{bao,fastgres} the existing query optimizer. The system model for this interaction is shown in Figure~\ref{fig:sysmod}. 

% \zi{Maybe summarize that LimeQO sits outside the DBMS, tracks actual execution time, and interacts with the query optimizer?  Then go into detail?  And maybe we motivate that hint-sets are the way the external module can still force the optimizer to explorew different plans in the space?}

Our prototype system was implemented to target the query optimizer for PostgreSQL, which features one of the most complex query optimizers in a widely used open source query engine. Numerous systems, including Amazon RedShift, TimescaleDB, and EnterpriseDB all build upon PostgreSQL's core, and our system would naturally extend to them. Additionally, the general approach of query hints is used by several other query optimizers, such as those in Presto~\cite{presto_meta}, MySQL, and Microsoft's SQL Server and SCOPE~\cite{bao_scope2}.%  Thus, our approach could likely be adapted to these and other systems.

%\zi{A bit about generality: What optimizers/systems would our system be directly applicable to? e.g., besides PostgreSQL (open source most complex qo), Redshift, TimescaleDB, EnterpriseDB, almost certainly.  What systems could it potentially be extended to, e.g., SQL server? Presto~\cite{}, Scope ~\cite{}}

Our framework has two paths: an online path, in which user-submitted queries are executed using plans that have been verified to be fast, and an offline path, where \lime can perform offline exploration. In the \emph{online path}, \textcolor{red}{\circleOne}~user-submitted queries are received by the DBMS' traditional optimizer. Then, \textcolor{red}{\circleTwo}~the optimizer asks \lime if a better query plan has been observed for this query. \textcolor{red}{\circleThree}~\lime replies with either a query plan that is faster than the default plan, or the default plan. \textcolor{red}{\circleFour}~This verified plan is then executed, \textcolor{red}{\circleFive}~and the results are returned. In the \emph{offline path}, \lime searches for better query plans. This offline search could happen when the DBMS is idle~\cite{resource_predict}, or could be performed on a snapshot of the database. During this time, \lime will \circleOne~predict the performance of all query plans in the workload matrix, and then \circleTwo~select the most promising query plans to explore. Next, \circleThree~these promising plans are executed, using a timeout based on the current best-known plan for that query. Once the new query plan finishes executing or times out, \circleFour~the performance is recorded. \circleFive~The newly observed values are finally stored back into the workload matrix. Importantly, the workload matrix contains two types of entires: (a) \emph{complete} entries, representing query plans whose latency was observed via execution, or \emph{censored} entries, representing queries plans that timed out, but for which a lower bound on their execution time is now known (e.g.,~if a query plan times out after 2 minutes, we assume that the true latency of that plan is greater than 2 minutes).

%\rcm{we should expand the above to include all the assumptions we are making: repetitive queries that we can match exactly, that queries have mostly-consistent performance, etc.}

\sparagraph{Our goal.} A naive implementation of \lime could simply evaluate random unobserved query plans (i.e., test blank entries in the workload matrix), but this strategy could waste offline execution time testing bad plans. Executing the entire matrix exhaustively is impractical; for example, processing the full CEB~\cite{flowloss} workload would take 12 days, and the Stack~\cite{flowloss} workload would require even longer than 16 days. Thus, \lime must strategically use each moment of offline execution time to create the largest improvement to the overall workload. \\

\sparagraph{Assumptions.} We make the following assumptions:
\begin{enumerate} [topsep=3pt,leftmargin = 15pt]
    \item {The DBMS repeatedly executes a set of queries, referred to as the workload.  Identical queries occur multiple times, and we cache them after the second occurrence and begin optimizing their performance. \new{Thus, the workload consists of a set of unique queries.}}
    \item {Each query in the workload has a default plan chosen by the underlying query optimizer, as well as a set of alternative plans, or hints ~\cite{bao, autosteer, fastgres}, a commonly used technique to guide query optimization.}
    \item {Queries generally exhibit consistent performance, meaning that the measurements obtained during the offline phase closely mirror those during online execution. Some may argue that data shifts might occur between the offline and online stages, potentially affecting performance. To address this concern, we provide two key observations. First, we demonstrate that near-optimal performance can be achieved after a duration equivalent to the total workload time (Section  ~\ref{sec:expr_perf}), making the time gap between the two phases negligible. Second, we show that even if data shifts occur, the best hint typically remains the same (Section ~\ref{sec:expr_data}), thus not significantly impacting our performance gains.}
\end{enumerate}

% \smallskip

\sparagraph{Why target repetitive workloads?} At first glance, considering only \emph{repeating} queries may seem like a major restriction. While there are certainly workloads with few or no repeating queries, there are also workloads like live dashboards that are almost purely repetitive~\cite{autosteer}. Recent studies of the AWS Redshift analytics database product found that more than 50\% of the queries executed on the Redshift fleet were repeated within 24 hours~\cite{holon, predicate_caching,  stage}, 75\% are repeated within a week, and 80\% within a month. Furthermore, analyses have shown that long-running queries (those taking longer than 1 hour) almost always repeat across all clusters~\cite{redset}. In Microsoft's SCOPE database, over 60\% of the job volume is recurring~\cite{bao_scope2}. Thus, targeting repeated workloads is both practical and impactful.

\sparagraph{Handling novel queries.} However, focusing solely on repetitive queries is still a major limitation to practitioners: new queries might not come along very often, but new queries almost certainly are introduced over a long period of time. Thus, we can say that new queries are \emph{rare but guaranteed}. \lime can support such rarely-arriving new queries in an intuitive way: we simply add new rows to the workload matrix. The prior entries in the workload matrix can potentially help predict the entries for the newly added rows (queries). The first time a new query is added, it is always executed using the underlying DBMS' default plan to avoid regressions, so one cell of the new rows is initialized. We evaluate and discuss \lime's performance on novel queries in Section~\ref{sec:expr_workload}.

\sparagraph{Why use query steering / hints?} We  assume, like prior work~\cite{autosteer, bao, fastgres, bao_scope, bao_scope2}, that the underlying DBMS' query optimizer supports ``hints'' that change the behavior of the optimizer. Each hint is a coarse-grained knob in the optimizer that impacts query selection, for example, disabling or enabling a particular join operator. For simplicity, we use ``hint'' to refer to a specific configuration of the optimizer, which in some systems may mean a combination of multiple distinct hints (i.e., ``hint sets''~\cite{bao}).

Casting learned query optimization as a hint selection problem, first done \new{by Bao} ~\cite{bao}, as opposed to fully replacing a query optimizer with a learned component~\cite{neo,balsa,lero}, or deeply integrating a solution with a particular optimizer, has a number of advantages. First, the ``hinting'' interface as described is implemented by a wide variety of databases, including PostgreSQL~\cite{url-pg_hints}, Presto~\cite{presto_meta}, SCOPE~\cite{bao_scope}, and RedShift~\cite{redshift}. Second, while optimizer hints are coarse-grained, they are also robust: selecting a plan that has been generated by a traditional query optimizer with a particular hint is much more likely to result in a reasonable plan than finer-grained techniques~\cite{bao}. Third, hints have \emph{enough} granularity to significantly improve a wide variety of analytic queries~\cite{bao_scope}, making hint steering an especially good match for repetitive analytic workloads. 

%\rcm{should update tthis to cite the new RedSet paper as well as Kipf's predicate caching paper, as well as Pavlo's Holon paper}

\section{Offline Exploration} 
\label{sec:offline_explore}

%\rcm{We first begin by formulating the problem as \emph{active learning over a low rank matrix} (or something like that). Then, we explain how this problem can be solved using linear or deep learning techniques.}

In this section, we first formulate the offline exploration problem (Section ~\ref{sec:prob-def}). Then we introduce our active learning exploration policy (Section ~\ref{sec:active-learning}), and extend it with two predictive models: a linear method (Section ~\ref{sec:linear}) and a neural method (Section ~\ref{sec:neural}).

\subsection{Problem Definition}
\label{sec:prob-def}

\sparagraph{Formulation.} %\rcm{need this to include timeouts / censored observations as well}
Let $Q = \{q_1, \ldots, q_n\}$ be a set of regularly executed queries, and let $H = \{h_1, \ldots, h_k\}$ be a set of hints. We define a workload matrix $\W$ as a $n \times k$ matrix that holds the performance metric (e.g., latency) for each query (row) and for each hint (column): that is, $\W_{ij}$ represents the latency of running query $q_i$ with hint $h_j$. 
\new{Since exactly computing $\W$ is prohibitive (i.e., requiring $nk$ query executions)}, we assume we only have access to a \emph{partially observed} copy of $\W$, denoted as $\WObserve$:
\begin{align}
    \WObserve_{ij} = 
    \begin{cases}
    \W_{ij} & \text{if } \W_{ij} \text{ is observed} \\
    \infty & \text{otherwise}
\end{cases}
\end{align}
When a query $q_i \in Q$ arrives, we select the hint $h_j$ with the best observed latency, that is, the minimum value in the row $\WObserve_i$. 

Our goal is to design an \emph{exploration policy} to reveal unobserved entries that can optimize performance while minimizing the offline time spent revealing entries of $\WObserve$.
% We call the set of queries $\{q_1, \ldots, q_n\}$ that need to be executed on a regular basis workload $Q$. We model each hint set $\h \in H$, the family of hint sets. During query execution, a hint set is passed to the query optimizer, and the query is converted to a plan tree and executed, producing some latency. Therefore, $\h : Q \rightarrow \R $ maps a query  $q\in Q$ to latency $t$. 
% The workload matrix $\W$ holds the latency of queries with different hint sets. As shown in Fig. \ref{fig:ex-matrix}, $\W_{i,j}$ represents the latency of running query $q_i$ with hint set $h_j$. 
% If all entries of $\W$ are known, the optimal workload latency $T(\W)$ can be achieved by 
% \begin{align}
%     T(\W) = \sum_i \min(\W_i)
% \end{align}
% However, observing the whole matrix is prohibitively expensive. For example, Completing the CEB workload used in experiments in Section X takes more than two months.
% the matrix on  and evaluating suboptimal plans incurs significant time costs. Instead, we have a partially observed matrix $\WObserve$ where
% \begin{align}
%     \WObserve_{i,j} = 
%     \begin{cases}
%     \W_{i,j} & \text{if } \W_{i,j} \text{ is observed} \\
%     \infty & \text{else}
% \end{cases}
% \end{align}
We define $P$ as the current workload latency,\footnote{Practitioners may also be interested in optimizing tail latency instead of total latency, in which case $P$ can be defined as the tail latency of the workload.} (i.e., the sum of the minimum observed values for each query) as follows:
\begin{equation}
    P(\WObserve) = \sum_{i=1}^{n} \min_{1 \leq j \leq k} \WObserve_{ij}
\end{equation}

\noindent and we define $T$ as the offline exploration  time required for revealing entries in the matrix to attain $\WObserve$:
\begin{equation}
    T(\WObserve) = \sum_{i=1}^{n} \sum_{j=1}^{k} \WObserve_{ij} \cdot \mathbf{1}_{\{\WObserve_{ij} \neq \infty\}}
\end{equation}
 
% We want to propose an iterative update algorithm to reduce the latency $P(\WObserve)$ by offline executing a set of $m$ query-hint pairs $p = \qhp$ and filling in the corresponding unobserved entries in each iteration such that the total workload latency will decrease. 
The challenge lies in minimizing both the workload latency $P(\WObserve)$ and the total offline exploration time $T(\WObserve)$ simultaneously. While $P(\WObserve)$ can be minimized by fully exploring all entries of $\W$, this would maximize $T(\WObserve)$. Conversely, $T(\WObserve)$ can be trivially independently minimized by doing no exploration at all, leading to sub-optimal $P(\WObserve)$ performance. Thus, we seek an algorithm to achieve both objectives simultaneously. 
%It operates by picking a set of $m$ most-promising query-hint pairs $\qhp$ during offline processing. By filling in the corresponding unobserved entries in $\WObserve$, it progressively reduces the total workload latency. The intuition is that if we observe a plan with lower latency than the current best plan, overall workload latency will decrease. 
% At step $k$, The offline exploration time $\TExplore$ is updated by
% \begin{align}
%     \TExplore^k \leftarrow \TExplore^{k-1} + \sum_{i}{\W_{p_i}}
% \end{align}

% \sparagraph{Ranking policy} 
% We will start $\WObserve$ with the PG default plans filled in and use a policy $P$ to pick the most promising unobserved values at each iteration.
% The policy $P: \WObserve \rightarrow \qhp$ will rank the unobserved values by their potential and return the top-m pairs with the greatest potential. We want to find the policy that can minimize $T(\WObserve)$ with the minimum $\TExplore$. 

\smallskip

\sparagraph{Timeouts.} A key aspect of the offline exploration process is that we are only interested in hints that outperform the current best observed hint. Therefore, we can safely optimize exploration by introducing a timeout limit, $\T_{ij}$, for each entry in the matrix, which is set to the current minimum latency observed in the corresponding row of $\WObserve$:

\begin{equation} \T_{ij} = \min_{1 \leq j \leq k} \WObserve_{ij} \end{equation}

This allows us to update the workload matrix $\WObserve$ by applying the timeout condition: 

\begin{align} 
\WObserve_{ij} = 
\begin{cases} \W_{ij} & \text{if } \W_{ij} \text{ is observed} \\ 
\T_{ij} & \text{if } \W_{ij} \text{ exceeds the timeout} \\
\infty & \text{otherwise} 
\end{cases} 
\end{align}

\noindent
\new{We note that $\WObserve$ evolves dynamically during the execution of the algorithm, as new entries in the matrix are observed.}

By incorporating timeouts, we can bound the time spent on exploration of hints that will show no performance improvement -- even if our predicted exploration time was incorrect.

\begin{algorithm}[t]
\caption{\lime}
\label{alg:offline-mc}
\LinesNumbered
\KwIn{$\WObserve$: initial observed matrix; $\Mask$: mask matrix; $\T$: timeout matrix; $pred$: predictive model}
\KwOut{Hint selections $[h_1,\ldots,h_n]$ for workload}
	\While{$\Mask \neq \mathbf{1}$ {\label{alg:while}}}
	{
	$\hat{\W} \leftarrow pred(\WObserve, \Mask, \T) $\label{alg:pred-step}\;
     \For{$i = 1$ to $n$ {\label{alg:it}}}
     {
     $h_j \leftarrow H[\text{argmin}_j(\hat{\W}_{ij})$]\; 
     $r{i} \leftarrow (\min{\WObserve_i} - \hat{\W}_{ij}) / \hat{\W}_{ij}$\; 
      add $ (q_i, h_j) $ to $S$ if $r_i > 0$ {\label{alg:compute-improve}}\;
     }
        $\text{Select top $m$ largest } (q_i, h_j) \text{ from } S \text{ w.r.t. } r_i$\; \label{alg:select-largest}
        \If{not enough to select {\label{alg:not-enough}}}
        {
        randomly select some unobserved $(q_i, h_j)$\; \label{alg:random-select}
        }
        $\T_{ij} = \min(\min(\WObserve_i), \hat{\W_{ij}} \times \alpha)$\;
        Offline execute, timeout if $\W_{ij} \geq \T_{ij}$\; \label{alg:timeout}
        Update $\Mask$, $\T$, and $\WObserve$\;} \label{alg:add}
        
    \For{$i = 1$ to $n$ {\label{alg:it-workload}}} 
        {$h_i \leftarrow H[\text{argmin}_{j} (\WObserve_{ij})]$\;} \label{alg:select-hint}
     \Return{$[h_1,\ldots,h_n]$}
\end{algorithm}

\subsection{Active Learning on a Low-Rank Matrix}
\label{sec:active-learning}

 Active learning strategies can be employed to efficiently explore and reveal unobserved entries in the matrix. Instead of exhaustively probing all entries, an active learning approach aims to identify the most informative entries to query, which can significantly reduce the exploration cost.

Here we propose two active learning techniques: \greedy and \lime (Algorithm ~\ref{alg:offline-mc}).

\smallskip
\sparagraph{Greedy.} Greedy does not rely on any predictive model. It selects the queries with the largest current minimum observed latency, ie.  argmax$_{i}(\min_{1 \leq j \leq k} \WObserve_{ij})$. Then for each query, we randomly select an unobserved hint. This strategy focuses on improving queries with the worst observed performance, as they offer the greatest potential for reducing the overall workload latency $P(\WObserve)$.

The underlying assumption of the greedy technique is that there is a correlation between the \emph{duration of a query plan} and that query plan's potential \emph{room for improvement}. While this assumption is often true in academic benchmarks (where we select long-running queries precisely because they have a lot of room for improvement --- otherwise the benchmark would not be very interesting), this assumption might not be true in practice. For example, the longest-running queries on Amazon RedShift are normally \texttt{COPY} queries or ETL jobs~\cite{redshift_workload} (e.g., a query that dumps the result of a simple scan to a CSV file). These types of queries have almost no room for improvement, since they are almost entirely bounded by write speed. We demonstrate this experimentally in Section~\ref{sec:expr_perf}. Nevertheless, we evaluate Greedy as a useful baseline.

\medskip

\sparagraph{\lime.}
The \lime approach, on the other hand, uses a predictive model to guide exploration. We will first use the predictive model to complete the partially observed matrix $\WObserve$ and generate the predicted matrix $\WPredict$\new{, which fills in the unobserved entries using estimated values}. Then, \lime selects query plans with the largest expected benefit, balancing the minimization of both $P(\WObserve)$ and the offline exploration time $T(\WObserve)$. For each query $q_i$, we compute the expected improvement ratio as follows:

\begin{equation}
    r_{i} = \left(\min_{1 \leq j \leq k} \WObserve_{ij} - \min_{1 \leq j \leq k}\WPredict_{ij}\right) / \min_{1 \leq j \leq k} \WPredict_{ij}
\label{eq:ratio}
\end{equation}
\vspace{0.5em}

This ratio captures the potential performance improvement from exploring the predicted best hint for query $q_i$ compared to its current best observed hint. By normalizing with the predicted best latency, we ensure that exploration focuses on minimizing both $P(\WObserve)$ and $T(\WObserve)$.

Algorithm ~\ref{alg:offline-mc} presents \lime in detail: Given an initial $\WObserve$, we use a predictive model to construct an estimate $\hat{\W}$ (Line~\ref{alg:pred-step}). With the estimated value, we go through every row of the predicted matrix and compute \emph{expected improvement ratio} (Equation ~\ref{eq:ratio} and Line ~\ref{alg:compute-improve}).
%We will also make sure only predicted improvements are considered (Line~\ref{alg:it}-\ref{alg:compute-improve}). 
The top $m$ queries (Line~\ref{alg:select-largest}), based on this improvement ratio, are then selected for exploration. In the case where there are fewer than $m$ positive predicted improvements (Line~\ref{alg:not-enough}), we will randomly select some unobserved entries (Line~\ref{alg:random-select}) to observe. Finally, we execute the $m$ selected plans, timeout the plan if the plan's latency is greater than $\T_{ij}$ (Line ~\ref{alg:timeout}), record their latency, and update $\Mask$, $\T$, and $\WObserve$ (Line~\ref{alg:add}). This process can be repeated until there is no more offline exploration time left, or when the algorithm stops finding potential improvements. Finally, we will return the current best hint for each query \mbox{(Line~\ref{alg:it-workload}-\ref{alg:select-hint})}. 

\sparagraph{Why not use existing active learning approaches?}
Our two proposed methods join an extensive literature of methods within the active learning community~\cite{settles2009active,deep_active_learning_survey}. Surprisingly, many existing active learning techniques make fundamental assumptions that render them unsuitable for our environment.  For example, most of the techniques specific to active matrix completion assume that the cost of acquiring the label is fixed~\cite{mc_queries,active_mc,activelearning_search}, and thus there is no need to balance the potential improvement in query time versus the cost of exploring a query hint. 
\new{Additionally, the scoring function in the Bayesian active learning method ~\cite{silva2012active} are less suitable for the unique characteristics of our problem space, since in our problem the cost of revealing a matrix cell is correlated with the matrix cell’s value (i.e., the cost of revealing the plan is the plan latency). Thus, we intentionally designed our scoring function to prioritize the efficient discovery of optimal hints for query optimization.}
Even methods such as selective supervision~\cite{selective_supervision}, which balance cost against improvement, assume a uniform cost within each class. The few methods we found that do not assume a fixed value or fixed cost were based on multi-armed bandits ~\cite{fair_active_learning, active_learning_pool, cmab_active_learning, active_learning_as_mab}, which assume that the pool of unlabeled observations is so large that taking an unlabeled 
%\old{observations} 
\new{observation} and labeling it does not change the distribution of unlabeled observations. This is patently untrue in our scenario, since observing a matrix entry obviously prevents that same matrix entry from being observed again. We tested each of these approaches to see if they would work even if their core assumptions were violated, but we were not successful. %Thus, we empirically found that our methods, which were much more focused on cost and benefit, were significantly more effective.

%\textcolor{orange}{TBD: \sparagraph{Why not use other active learning techniques?} Active learning is its own research community and problem area, with a wide variety of techniques being developed over several decades (see~\cite{settles2009active,deep_active_learning_survey} for surveys). Unfortunately, there are several reasons why simply lifting an active learning technique from the literature is not easy. First, many of the most successful active learning techniques, like X or Y, assume that the \emph{cost} of acquiring a piece of information is not correlated with the \emph{value} of that particular information (e.g.,} 

%\smallskip

%\zi{Should this paragraph below be promoted to Section 3.3, and the methods 3.3.1 (Linear) and 3.3.2 (Neural)?}

\subsection{Predictive Model}
In real-world scenarios, the workload matrix $\W$ often exhibits low-rank structure due to inherent correlations among queries and hints. Leveraging this property, we propose two predictive models: linear methods based on matrix completion~\cite{als} (Section~\ref{sec:linear}) and neural methods based on low rank embeddings and tree convolution neural networks (TCNN)~\cite{tree_conv,neo} (Section~\ref{sec:neural}). These two models can be easily integrated as the predictive model in Algorithm ~\ref{alg:offline-mc}.

\subsubsection{Linear Method}
\label{sec:linear}

\begin{algorithm}[t]
\caption{ALS}
\label{alg:als}
\LinesNumbered
\KwIn{$\WObserve$: observed matrix; $\Mask$: mask matrix; $\T$: timeout matrix; $r$: rank; $\lambda$: regularization parameter; $t$: number of iterations}
\KwOut{Completed matrix $\hat{\W}$}
	Initialize $\A, \B$ of size n $\times$ r, and k $\times$ r randomly \;
	\For{$i = 1$ to $t$}
	{
	    $\hat{\W} \leftarrow \Mask \odot \WObserve + (1-\Mask) \odot \A\B^T$ \label{alg:update-1}
     
        \If{
        $\hat{\W} < \T \text{ and } \T > 0$
        } {$\hat{\W}  = \T$ {\Comment{Censored technique}}} \label{alg:censor-a}
        $\A \leftarrow \hat{\W} \B (\B^T\B +  \lambda I)^{-1}$ {\Comment{Update $\A$ with least squares solution}} \label{alg:compute-a}
        
        {$\A[\A < 0] = 0$ {\Comment{Ensure non-negative entries}} \label{alg:non-negative-a}}
        
        {$\hat{\W} \leftarrow \Mask \odot \WObserve + (1-\Mask) \odot \A\B^T$ \label{alg:update-2}}
        
         \If{
        $\hat{\W} < \T \text{ and } \T > 0$
        } {$\hat{\W}  = \T$ {\Comment{Censored technique}}  }\label{alg:censor-b}
        $\B \leftarrow \hat{\W} \A (\A^T\A +  \lambda I)^{-1}$ {\Comment{Update $\B$ with least squares solution}} \label{alg:compute-b} 
        
        $\B[\B < 0] = 0$ {\Comment{Ensure non-negative entries}} \label{alg:non-negative-b}
	}
    {$\hat{\W} \leftarrow \Mask \odot \WObserve + (1-\Mask) \odot \A\B^T$}
    
    \Return{$\hat{\W}$}
\end{algorithm}

We apply \emph{matrix completion}~\cite{als} as the linear method. By assuming that $\W$ has low rank, the observed entries of $\W$ can be used to predict the unobserved entries. Notably, this technique uses the partially observed matrix $\WObserve$ directly, and does not rely on any properties of the queries or their plans (e.g., cost estimates, plan structure, operators).

\smallskip
\sparagraph{Matrix completion.} Matrix completion (MC) is a technique used to recover unobserved entries in a low rank matrix~\cite{nuclear, candes_power_2009, als, srebro_maximum-margin_2004}. We define $\Mask$ as the \emph{mask matrix}, which has the same shape as~$\WObserve$:~$\Mask_{ij} = 0$ if $\WObserve_{ij} = \infty$ and $\Mask_{ij} = 1$ otherwise (that is, $\Mask$ is one for observed entries of $\WObserve$ and zero otherwise). Given a partially observed $\WObserve$, a rank constraint $r$, and a regularization parameter $\lambda$, we can build an estimate of $\W$ as $\hat{\W} = \A \B^T$ by solving:
\begin{align}
\label{eq:als_obj}
    {\min_{\A,\B}}\hspace{0.1cm} \left[ \|\Mask \odot (\WObserve - \A\B^T)\|_{F}^{2} + \lambda \left( \|\A\|_{F}^{2} + \|\B\|_{F}^{2} \right)\right]
\end{align}
where $\A$ and $\B$ are $n \times r$ and $k \times r$ matrices, respectively, and $\odot$ represents the element-wise product. To find $\A$ and $\B$, we use the Alternating Least Squares (ALS) algorithm~\cite{als}, which is based on the following key observation: while Equation~\ref{eq:als_obj} is not \new{jointly} convex in $\A$ and $\B$, Equation~\ref{eq:als_obj} is convex in $\A$ for fixed $\B$, and vice versa. 
\new{Convexity in this context implies that optimizing $\A$ or $\B$ while the other is fixed ensures convergence to a global minimum for that subproblem. This property is crucial to the ALS algorithm, which alternatives between solving each convex subproblem, progressively improving the overall solution (although there is no guarantee that the final approximation is optimal).}
%\old{Thus, ALS works by alternating between fitting $\A$ assuming a fixed $\B$, and then fitting $\B$ assuming a fixed $\A$. }

Algorithm \ref{alg:als} details our modified version of the ALS algorithm: 
%Given the partially observed matrix $\WObserve$ and parameters including rank $r$, regularization parameter $\lambda$, and the number of iterations $t$, 
we iteratively update the matrices $\A$ and $\B$ using the least squares solution (lines ~\ref{alg:compute-a} and ~\ref{alg:compute-b}), and fill in the predicted entries (lines ~\ref{alg:update-1} and ~\ref{alg:update-2}).
Additionally, since query latencies are strictly positive, we impose non-negative constraints on $\A$ and $\B$ after each iteration (lines ~\ref{alg:non-negative-a} and \ref{alg:non-negative-b}).
\new{While this constraint may slightly reduce approximation flexibility, it ensures the data remains physically meaningful (i.e., positive), allowing the score function (Equation~\ref{eq:ratio}) to operate effectively. The non-negative constraint can be interpreted as a heavy-handed\footnote{\new{We say ``heavy-handed'' because negative entries in $\A$ or $\B$ do not necessarily result in negative predicted latencies.}} prior that query latency must be positive.}

\begin{figure}[t]
    \begin{minipage}{0.485\textwidth}
        \includegraphics[width=1\linewidth]{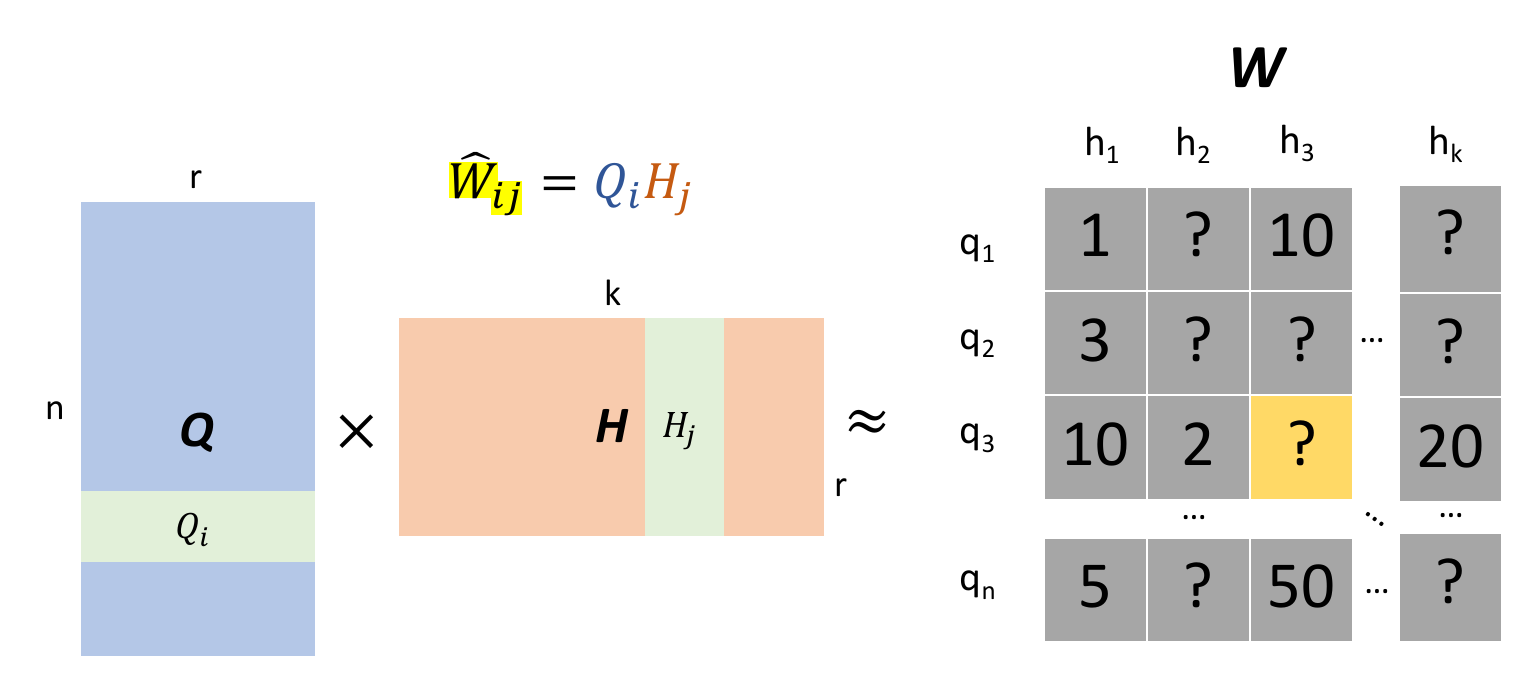}
        \vspace{1ex}
  \caption{Linear Method}
  \vspace{4ex}
        \label{fig:als}
    \end{minipage}
    \hfill
    \begin{minipage}{0.47\textwidth}
        \includegraphics[width=1\linewidth]{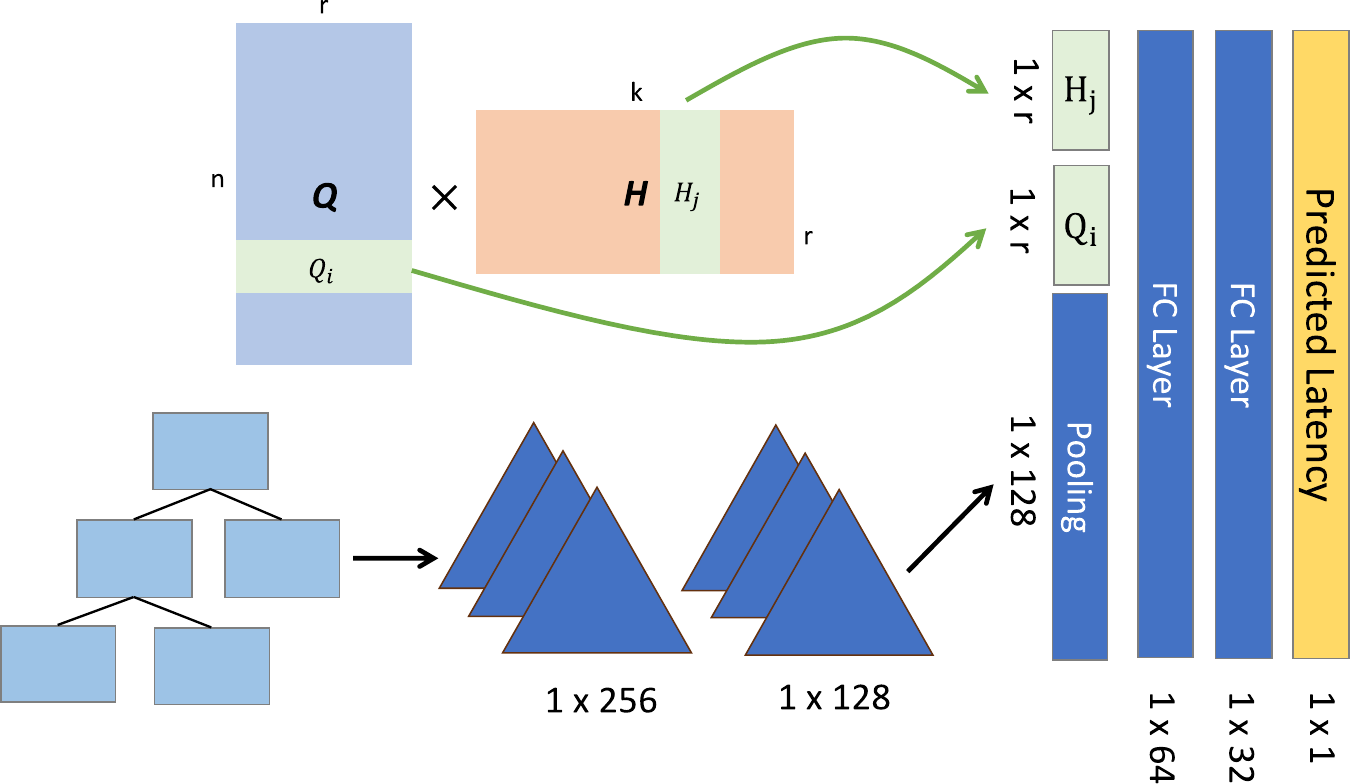}
        \vspace{-5ex}
  \caption{Neural Method}
        \label{fig:neural}
    \end{minipage}
\end{figure}

\sparagraph{Predicted Latency.} 
We illustrate the matrix factorization process in Figure~\ref{fig:als}. To calculate an unobserved entry, we simply compute the dot product of the corresponding two vectors: $\WPredict_{ij} = \A_{i} \B_{j}$. Thus, each row of $\A$ represents a ``query vector'' that contains information about the query in a particular row, and each column of $\B$ represents a ``hint vector'' that contains information about the hint in a particular column. We pick $\A$ and $\B$ such that the dot product of a query vector and a hint vector predicts the latency of a given query under that specific hint.

\smallskip

\sparagraph{Timeouts / Censored Technique.} 
To handle the timeouts during the exploration process, we incorporate a censored technique specifically designed for this scenario. We define the timeout matrix $\T$, with the same shape as $\W$, where only the timeout values are filled in, and all other entries are zeros. As shown in Algorithm ~\ref{alg:als}, lines ~\ref{alg:censor-a} and ~\ref{alg:censor-b}, if the predicted value for a timed-out entry does not meet the timeout threshold (i.e., it is smaller than the timeout), we manually set it to the timeout value to reflect the observed limitations. This way, future iterations of ALS will never try to predict a value less than the timeout, but if a prior iteration of ALS predicts a value greater than the timeout, that value will be kept for future iterations.

%explanation of matrix completion technique, what it assumes, and how ALS works. Can include a diagram with the decomposition we learn.

\subsubsection{Neural Method} 
\label{sec:neural}

%\rcm{need to include explanation of the new embeddings + TCNN technique}

Here, we present an alternative approach for offline query optimization using a neural network. In this approach, we assume query plan features are available (e.g., cost and cardinality estimates), and that the underlying query optimizer generates tree-structured plans, such as those generated by PostgreSQL~\cite{url-postgres}. 
Leveraging these additional features may provide better accuracy than \lime, albeit with increased computational overhead.

Systems like Neo~\cite{neo}, Bao~\cite{bao}, and Balsa~\cite{balsa} use plan features to learn a value function that estimates the overall cost or latency of executing a query. Similarly, we propose a new predictive \emph{transductive TCNN model}, which combines tree-structured features with the low-rank property of the workload matrix. We will first describe Tree Convolution~\cite{tree_conv}, followed by an explanation of how TCNN Embedding integrates Tree Convolution with low-dimensional embeddings.

\smallskip

\sparagraph{Tree Convolution.} 
Tree convolution~\cite{tree_conv} is an adaption of traditional image convolution, designed for tree-structured data. It applies tree-shaped filters to query plans, identifying patterns related to query performance~\cite{neo}. %This process is structured in layers, with initial layers spotting simple node types or relationships, and deeper layers recognizing more complex node arrangements. 

We first binarize the query plan trees as described in \new{Bao} \cite{bao}, and encode each tree node into a vector that includes: 1) a one-hot encoding of the operator, 2) cost and cardinality information. After the final layer of tree convolution, dynamic pooling and fully connected layers are used to predict query performance. Tree Convolutions can effectively capture structural patterns in query plans and serves as an inductive bias for solving query optimization problems (i.e., the structure of the TCNN network is biased towards learning features that are useful for query optimization~\cite{neo}). This technique has been widely applied in query optimization ~\cite{bao,balsa,roq,autosteer}. 

\smallskip
\sparagraph{TCNN Embedding.}
In contrast to the linear model approach (Section ~\ref{sec:linear}), where two matrices are used to replace a neural model and predict performance, TCNN Embedding combines low-dimension matrix representations with plan tree features using a neural network architecture. The key component of this approach is the \textbf{Embedding} layer, which provides compact vector representation of queries and hints. 

As shown in Figure ~\ref{fig:neural}, each query and each hint is mapped to a vector of size $r$ via the two embedding layers. These vectors are learned in such a way that similar queries and hints are represented by similar vectors. After retrieving the embedding vectors, we concatenate them with the outputs from the tree convolution layer. The concatenated vector, containing both structural information from the query plan tree and low-rank embeddings, is then passed through fully connected layers for performance prediction. This hybrid approach combines the advantages of tree convolution and low-rank embeddings, resulting in better performance compared to using either method alone.

The reason that the transductive TCNN captures the low-rank structure of the workload matrix is that the learned embeddings, labeled $\A$ and $\B$ in Figure~\ref{fig:neural}, are isomorphic to the linear decomposition matricies $\A$ and $\B$: the embedding for a particular query represents features for the entire row of the matrix, and the embedding for a particular hint represents features for an entire column of the matrix. Since the same query embedding is used for every entry in a row of $\W$, and since the same hint embedding is used for every entry in a column of $\W$, the transductive TCNN can be said to have \emph{weight sharing}~\cite{shared_weights}.

\smallskip
\sparagraph{Predicted Latency.}
In each offline exploration step, we train a TCNN Embedding model to predict the unobserved values in $\WObserve$. Specifically, the model is trained using the observed entries of $\WObserve$, using features extracted from each query plan as well as the corresponding query and hint indices $(i, j)$. After training, the model performs inference to generate predicted latencies for the unobserved entries.  Consequently, $\hat{\W}$ consists of the actual latencies for the observed entries and the model's predictions for the unobserved ones. Furthermore, the model is initialized with the weights from the previous step, enabling it to build on prior learning.

\smallskip
\sparagraph{Timeouts / Censored Technique.} 
Recall that for some entries in the matrix, we timed out at a certain threshold $\tau$ and thus the values represent a lower bound of the actual execution time.
%\zi{Can we explain this more clearly in case the reader wasn't closely following, i.e., that for some entries in the matrix, we timed out at a limit and thus the values represent a lower bound?}
To effectively incorporate these censored observations into our neural network model, we introduce a new loss function specifically for the timed-out values in the neural network model training process: 

\begin{equation}
    \mathcal{L}(\hat{y}, y, \tau) = \frac{1}{n} \sum_{i=1}^{n} \mathbf{1}_{\{\hat{y}_i < \tau_i\}} \cdot \left(\hat{y}_i - y_i\right)^2
\label{eq:censored_loss}
\end{equation}

\noindent where $\hat{y}_i$ is the predicted value, $y_i$ is the true value, and $\tau_i$ is the timeout value for the ith entry. The term $\mathbf{1}_{\{\hat{y}_i < \tau_i\}}$ is an indicator function, which is 1 if predicted value is less than the threshold and 0 otherwise. 
This loss function ensures that only predictions less than the timeout threshold contribute to the loss calculation, penalizing the model for incorrect predictions where it is certain to miss the true value, while not penalizing it for predictions where the correctness of prediction is uncertain.

By replacing the standard Mean Squared Error (MSE) loss with this censored loss function, we enable the model to appropriately handle timeout observations and effectively learn from them.
\section{Experiments}
\label{sec:expr}

\new{In this section, we conduct a series of experiments to evaluate our proposed techniques: \lime, the linear method introduced in Section ~\ref{sec:linear}, and \limeplus, the neural method introduced in Section ~\ref{sec:neural}.}
Our experimental results seek to answer the following questions:
\begin{itemize}[topsep=0pt,leftmargin = 9pt]
    \item {How does \lime and \limeplus's performance compare to simple baselines and existing techniques? (Section~\ref{sec:expr_perf})} {And how much overhead do \lime and \limeplus entail? (Section~\ref{sec:expr_overhead})} 
    \item {How well does \lime handle workload shift (Section~\ref{sec:expr_workload}) and data shift}? (Section~\ref{sec:expr_data})
    \item {How reasonable is our low rank assumption, and how much does each component and optimization of \lime matter (ablation studies)? (Section~\ref{sec:expr_ablation})} 
\end{itemize}

% \zi{Might be worth a comment about the LimeQO implementation?}

\smallskip
\sparagraph{Experimental setup.} We evaluated \lime using PostgreSQL 16.1~\cite{url-postgres}. \lime uses the same 49 hints as Bao~\cite{bao}, which are based on six configuration parameters where we can enable or disable hash join, merge join, nested loop join, index scan, sequential scan, and index-only scan\footnote{It is not possible to turn off all join operators or turn off all scan operators, hence 49 hints instead of 64.}. Each (query, hint) pair is executed five times on an AMD Ryzen 5 3600 6-Core Processor, running Arch Linux 6.11.3. For the subsequent experiments, we selected the median runtime for each pair. Additionally, each technique's experiments were repeated five times, and we report the average runtime along with the standard deviation.

\begin{figure*}[t]
    \centering
    \begin{minipage}{0.245\textwidth}
        \includegraphics[width=\textwidth]{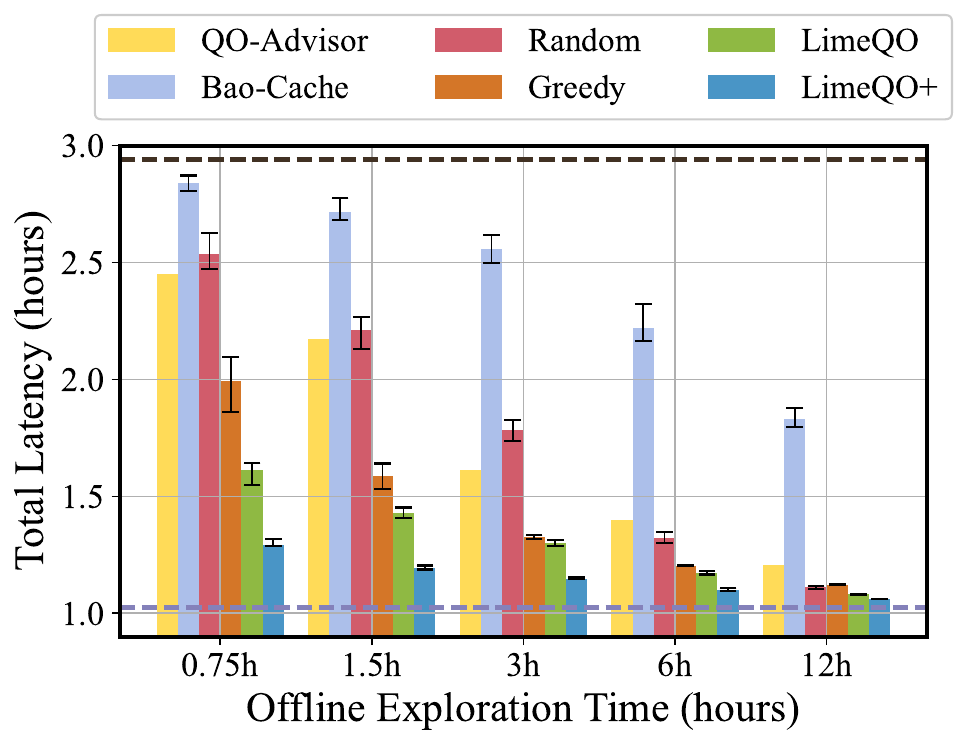}
    \end{minipage}
    \hfill
    \begin{minipage}{0.245\textwidth}
        \includegraphics[width=\textwidth]{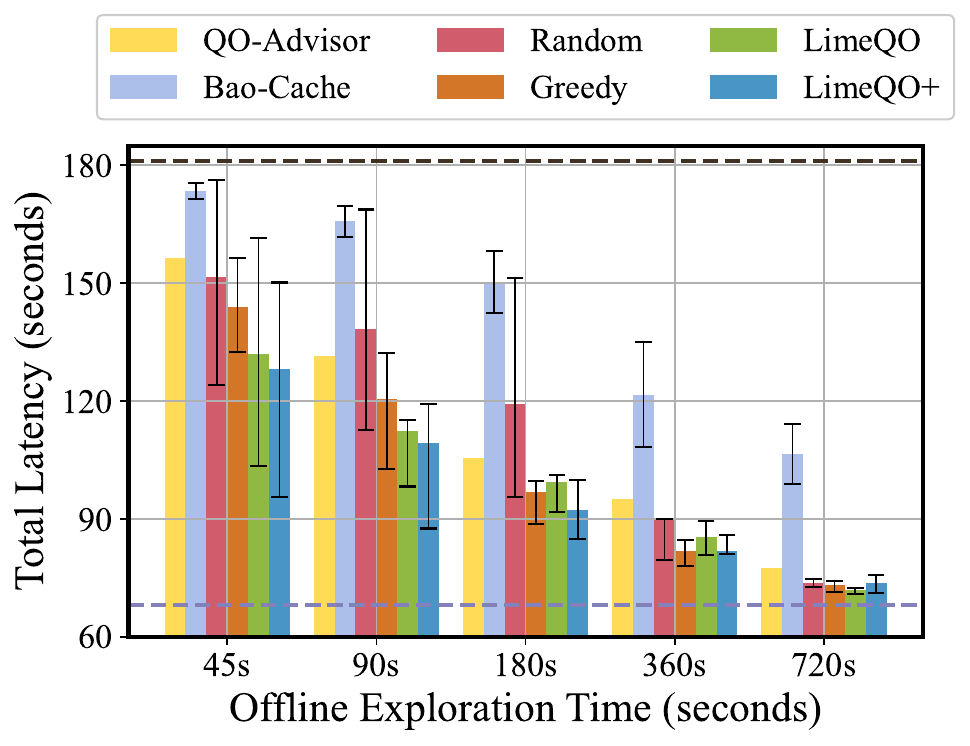}
        % \caption{Plot 2}
    \end{minipage}
    \hfill
    \begin{minipage}{0.245\textwidth}
        \includegraphics[width=\textwidth]{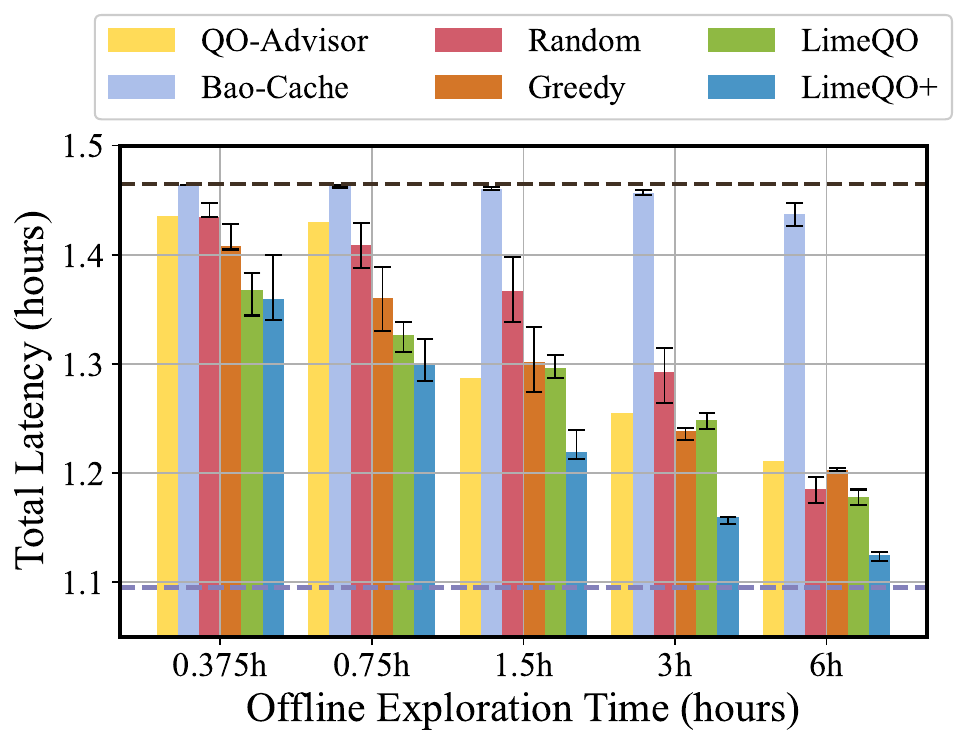}
        % \caption{Plot 3}
    \end{minipage}
    \hfill
    \begin{minipage}{0.245\textwidth}
        \includegraphics[width=\textwidth]{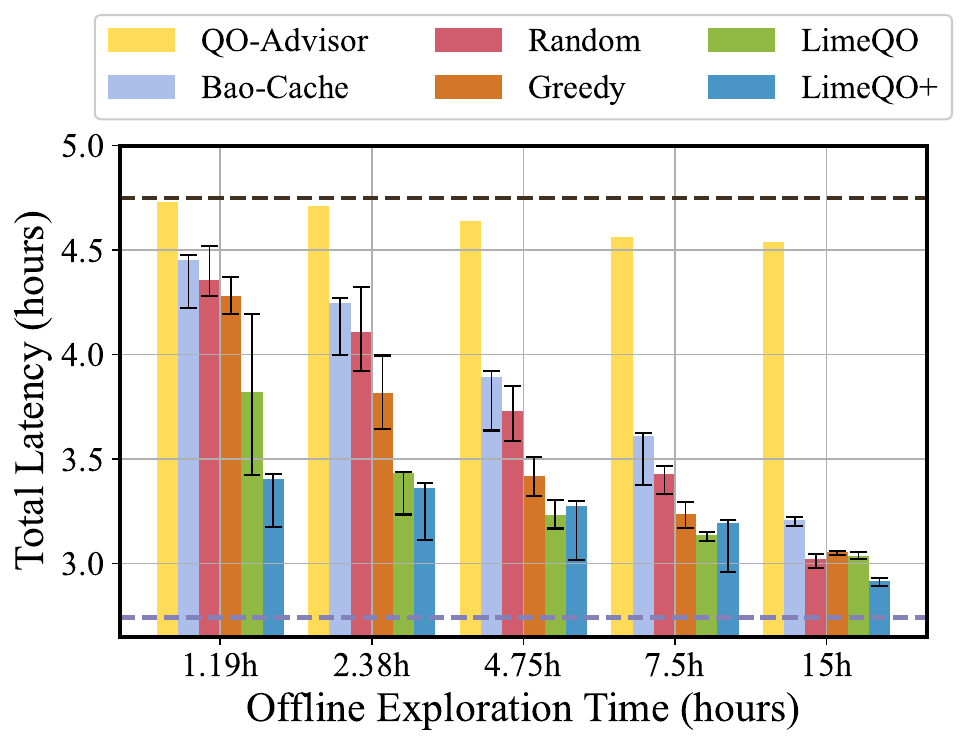}
    \end{minipage}
    \vspace{-1ex}
    \caption{\new{Performance Improvements measured on four different workloads: CEB, JOB, Stack, and DSB.} Goal is to achieve lower total latency with less offline exploration time. Offline exploration times chosen on the X-axis correspond to [1/4, 1/2, 1, 2, 4] $\times$ default workload time in each workload.}
    \label{fig:bar}
\end{figure*}

\begin{figure*}[t]
    \centering
    \begin{minipage}{0.32\textwidth}
        \includegraphics[width=\textwidth]{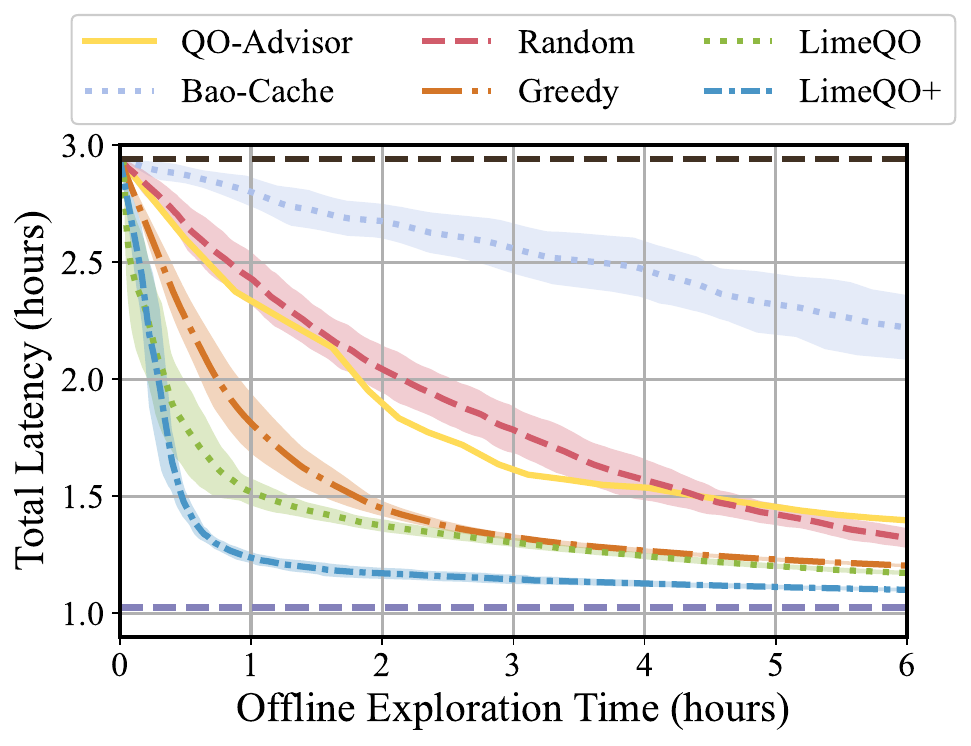}
        \caption{Total Latency vs. Offline Exploration Time on CEB workload. \lime and \limeplus outperform other techniques.}
        \label{fig:offline}
    \end{minipage}
    \hfill
    \begin{minipage}{0.32\textwidth}
        \includegraphics[width=\textwidth]{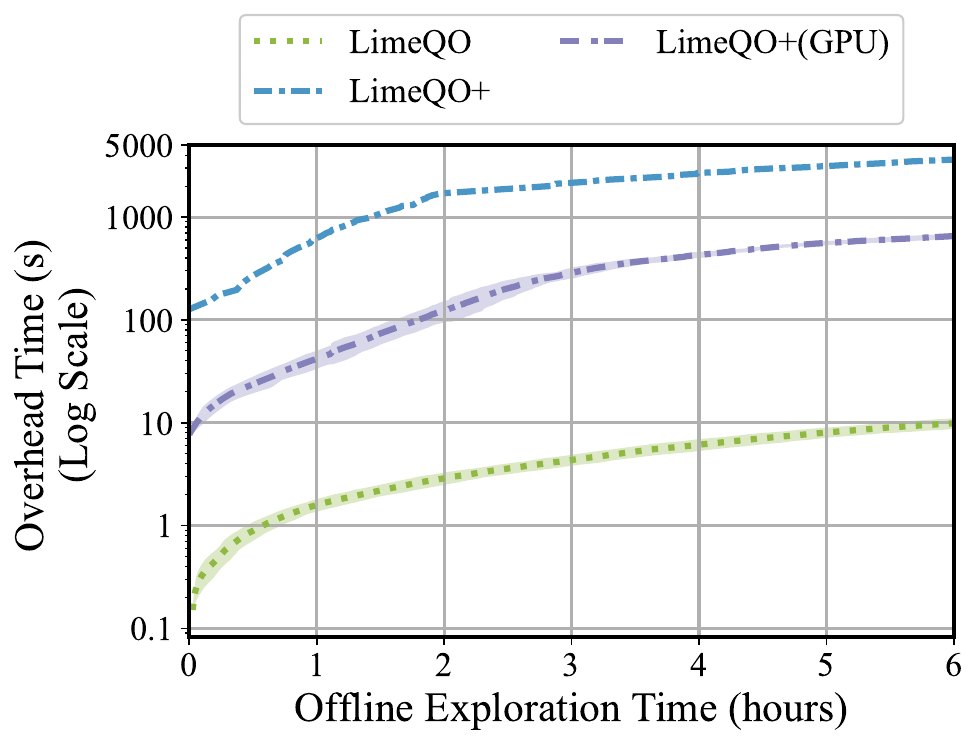}
        \caption{Overhead Time vs. Offline Exploration Time on CEB workload. On CPU, \limeplus spend 360x overhead time than \lime.}
        \label{fig:overhead}
    \end{minipage}
    \hfill
    \begin{minipage}{0.32\textwidth}
        \includegraphics[width=\textwidth]{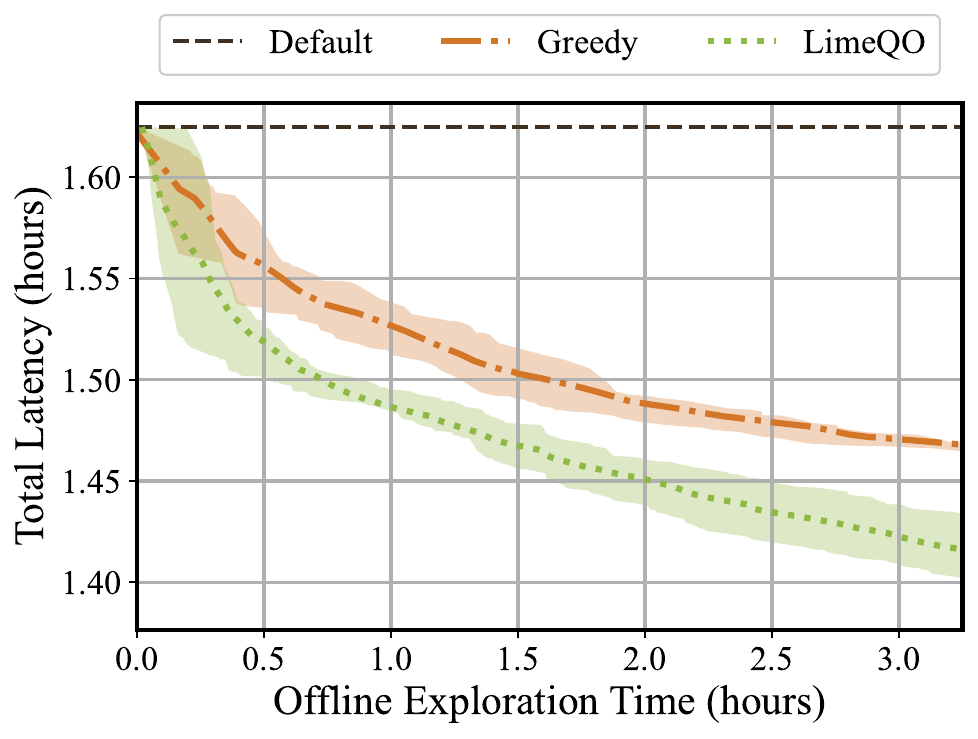}
        \vspace{-2.2em}
        \caption{Greedy vs. \lime after we add a ETL query into the Stack Workload. Note that the default workload time increased from 1.46 hours to 1.62 hours. %In this case, \lime consistently outperforms \greedy.
        }
        \label{fig:greedy}
    \end{minipage}
    %\vspace{-1ex}
\end{figure*}

% \begin{figure}[t]
%   \centering
%   %\vspace{3ex}
%   \includegraphics[width=1\linewidth]{fig/job_ratio.pdf}
%   \vspace{-2ex}
%   \caption{\new{Explored Queries Ratio on  JOB workload.}}
%   \vspace{-1ex}
%   \label{fig:job_ratio}
% \end{figure}

% \begin{figure}[t]
%   \centering
%   %\vspace{3ex}
%   \includegraphics[width=1\linewidth]{fig/ceb.pdf}
%   \vspace{-4ex}
%   \caption{Total Latency and Overhead Time vs. Offline Exploration Time on CEB workload. In addition to the bar plots in Figure \ref{fig:bar}, we are showing the trend of total latency reducing and overhead time accumulating over time.}
%   \vspace{-2ex}
%   \label{fig:offline}
% \end{figure}

  % \begin{minipage}[t]{0.235\textwidth}
  %   \centering
  %   \includegraphics[width=\textwidth]{fig/overhead.pdf}
  %   \vspace{-4ex}
  %   \caption{Overhead}
  %   \label{fig:overhead}
  %   \vspace{-3ex}
  % \end{minipage}

%\rcm{(1) how does limeqo's offline performance compare to other baselines? (2) same for online mode, (3) overhead, (4-X) other questions like rank}

\smallskip
\sparagraph{Workload and datasets.} 
We evaluated the performance of \lime using multiple diverse workloads from prior work, which are detailed in Table~\ref{tab:dataset}. 
% \zi{A sentence on each workload, what it represents, what makes it different from the others? In particular Stack is interesting because you have two snapshots, correct?}
The \textbf{JOB} workload~\cite{howgood} is relatively small, comprising only 113 queries on the IMDb database. The \textbf{CEB} workload~\cite{flowloss} expands upon JOB by adding thousands of additional queries. The \textbf{Stack} dataset contains over 18 million questions and answers from StackExchange websites (e.g., StackOverflow.com) collected over ten years. We have two snapshots of this data from 2017 and 2019. In Section~\ref{sec:expr_perf}, we use the 2019 version, and in Section~\ref{sec:expr_data}, we utilize both versions to model data shifts. 
\new{The \textbf{DSB} benchmark~\cite{dsb} is adapted from the TPC-DS benchmark~\cite{tpcds} with more complex data distribution and more varieties in queries. Specifically, we select a scale factor of 50 (following prior work~\cite{learned_shift, index_advisor_study}) and generate 20 distinct parameterized query instances from each query template.}

For each workload, we list the time it takes for a vanilla PostgreSQL database to execute every query (``Default''), and the theoretical \new{minimum} time achievable by a not-possible-in-practice oracle function (``Optimal''). Each workload has between 1.36x to 2.66x ``headroom'' (Default/Optimal).

%\begin{itemize} [topsep=0pt,leftmargin = 10pt]

%\item \textbf{CEB} (Cardinality Estimation Benchmark) ~\cite{flowloss} contains 3133 queries over the IMDb database.

%\item \textbf{JOB} (Join Order Benchmark) ~\cite{howgood} contains 131 queries over the IMDb database. This is a relatively small workload, which only took 181 seconds to complete by default.

%\item \textbf{Stack} ~\cite{flowloss} contains 6191 queries over the Stack database ~\cite{bao}. The Stack database is collected from the StackExchange websites.

%\end{itemize}

%All experiments were conducted on a server running 64-bit Ubuntu 22.04 with Intel(R) Xeon(R) Gold 6248R CPU @ 3.00GHz, 503 GB RAM, and an NVIDIA A100 GPU. Without \lime, PostgreSQL runs the CEB workload in 3 hours. Each experiment is executed five times; we plot the average and standard deviation.

\begin{table}[h]
%\vspace{1ex}
\begin{tabular}{@{}cccccc@{}}
\toprule
Workload & Dataset  & Size    & \# Queries & Default  & Optimal \\ \midrule
JOB ~\cite{howgood}           & IMDb    & 7.2 GB      & 113      & 181 s         &  68 s        \\
CEB ~\cite{flowloss}           & IMDb   & 7.2 GB      & 3133      & 2.94 hrs         &  1.02 hrs      \\
Stack \cite{bao} & Stack & 100 GB & 6191         &  1.46 hrs         &   1.09 hrs     \\ 
DSB ~\cite{dsb} & DSB & 50 GB & 1040         &  4.75 hrs         &   2.74 hrs \\
\bottomrule
\end{tabular}
\vspace{1ex}
\caption{Four workloads we covered in the experiments. Default refers to the total time taken with PostgreSQL's default hint, while Optimal is the best time achievable if all hints were explored.}
\label{tab:dataset}
\vspace{-6.5ex}
\end{table}
%\smallskip

\sparagraph{Techniques and tests.} We compare six different methods. For each method, we initially reveal the entries in the workload matrix corresponding to the default plan produced by PostgreSQL, simulating an environment where queries are executed repeatedly.

\begin{itemize} [topsep=0pt,leftmargin = 9pt]
\item{\new{\QOAdvisor: the QO-Advisor technique ~\cite{bao_scope2} adapted to PostgreSQL. Instead of using a contextual bandit model that learn from the estimated cost to recommend single-rule flips, we select the unexplored entry with the lowest optimizer cost (this is the best action that QO-Advisor's contextual bandit could possibly pick, since QO-Advisor's multi-armed bandit operated over the optimizer's cost model).}}
\item{\new{\BaoCache: the technique of Bao ~\cite{bao} adapted to offline exploration. The TCNN is used to select unobserved entries to explore. We cache the results and select the best observed hint for each query (thus guaranteeing that there are no query regressions).}}
\item{\rand: explore the workload matrix by randomly selecting unobserved entries.}
\item{\greedy: explore the matrix by selecting the longest running queries as described in Section ~\ref{sec:active-learning}, then randomly picking the unobserved hints.}
\item{\lime: use MC to explore the matrix as described in Section~\ref{sec:linear}. We set $r = 5$, $\lambda = 0.2$ and $t = 50$ in Algorithm \ref{alg:als}. We implemented it using standard linear algebra libraries, specifically NumPy's numpy.linalg which uses LAPACK ~\cite{lapack} at core.}
\item{\limeplus: use TCNN Embedding as described in Section ~\ref{sec:neural}. For the TCNN component, we use the same TCNN architecture as~\cite{bao}, except that we add a dropout layer~\cite{dropout} with $p=0.3$ between each tree convolution layer, which universally improved results. For the embedding layer, we set $r = 5$. Training is performed with Adam~\cite{adam} using a batch size of 32, and is run for 100 epochs or convergence (defined as a decrease in training loss of less than 1\% over 10 epochs) is reached.}
%\item{\tcnn: uses TCNN to predict unobserved matrix entries as described in Section~\ref{sec:neural}. We use the same TCNN architecture as~\cite{bao}, except that we add a dropout layer~\cite{dropout} with $p=0.3$ between each tree convolution layer, which universally improved results. Training is performed with Adam~\cite{adam} using a batch size of 32, and is run for 100 epochs or convergence (defined as a decrease in training loss of less than 1\% over 10 epochs) is reached.}

%\old{\item{\Bao: the technique of Bao ~\cite{bao} adapted to offline exploration. The TCNN is used to select unobserved entries to explore, but the TCNN model is fully trusted in the online path (and thus regressions are possible).}}

\end{itemize}

\subsection{Performance Improvements} 
\label{sec:expr_perf}

% \rcm{First say what the figures show. eg Figure~\ref{fig:bar} shows the total workload time achieved after spending a certain amount of time optimizing... then pick a case study: we can see that for workload X after Y time, \lime has reduced the total workload time from A to B. We can rule out random chance as a cause for this because... Finally, end the discussion of the figure with a \emph{emphasized statement about what should be taken away from the experiment}.}

\sparagraph{How much can \lime and \limeplus improve latency?} Figure ~\ref{fig:bar} shows the total workload time across different workloads after a certain amount of offline exploration time. 

On the \textbf{CEB} workload, after 1.5 hours (50\% of the default workload time), \lime brings the latency down by 50\%, from 2.94 hours to 1.45 hours --- within 15\% of the optimal reduction (65\%). \limeplus achieves a 60\% reduction, lowering the time from 2.94 hours to 1.2 hours, just 5\% above the optimal. The relatively poor performance of \rand and \greedy indicates that these improvements are not due to chance. 
For the \textbf{JOB} workload, after one default workload time, \lime and \limeplus reduce processing time to 100s (a 45\% reduction) and 80s (a 56\% reduction) respectively, whereas the optimal is 68s (a 62\% reduction). 
\new{Figure~\ref{fig:job_ratio} further illustrates that \lime and \limeplus explored fewer queries over the offline exploration period. This suggests that our techniques prioritize filling in the ``more important" entries in the matrix rather than performing exhaustive search.}
On the \textbf{Stack} workload, after one default workload time, \lime and \limeplus reduce latency by 11\% and 17\% respectively, which is close to the optimal possible improvement of 25\%.
\new{For the \textbf{CEB} workload, after one default workload time, \lime and \limeplus reduce latency to 3.25 hours (a 32\% reduction),  and 3.26 hours (a 31\% reduction) respectively, whereas the optimal latency is 2.74 hours (a 42\% reduction).
}

We observe that, across different workloads, \limeplus can achieve better results than \lime in many cases, but this comes at the cost of additional overhead (further investigated in Section ~\ref{sec:expr_overhead}). Both \lime and \limeplus outperforms \rand and \greedy techniques at the start (0 to 1 $\times$ default workload time), although the four techniques converge after the 4 $\times$ default workload time. 
\new{Compared to \QOAdvisor and \BaoCache, our techniques consistently demonstrated better performance across various exploration durations and workload settings, highlighting the importance of considering the entire workload at once when making exploration decisions.}

%In conclusion, both \lime and \limeplus can significantly reduce workload latency with less offline exploration time than it takes to execute the workload once. 

% (description)Figure X shows blah blah blah.  (example) notably, after A hours of offline optimization, \lime brings the latency down by z\% from x minutes to y minutes, which is only q\% more than the optimal. (analysis) we find that \limeplus can achieve better results than \lime in many cases, but at the cost of additional overhead (investigated in REF) (signoff/emphasize) Both LimeQO and LimeQO+ can significantly reduce workload latency with less offline exploration time than it takes to execute the query once normally.

\smallskip

% \rcm{Then, start the next figure. We take a deeper look at workload Z in Figure~\ref{fig:offline}a (left hand side), seeing how workload time changes with time spent on optimization. Here, we additionally compare with Bao, and you can see regressions, etc.}

\sparagraph{Performance over time.}
We further analyze the CEB workload in Figure ~\ref{fig:offline}, where we show how different techniques' workload time changes with time spent on offline optimization.

Initially, \lime reduces workload latency more rapidly than \limeplus. However, after approximately 20 minutes of exploration, \limeplus surpasses \lime, achieving a lower total latency. This shift can be attributed to \limeplus's deep learning approach, which improves its performance as it receives more training data. 
% However, this advantage comes at the cost of much larger training, inference time, and space overhead. Additionally, \limeplus requires rich plan tree features and feature extraction, while \lime does not require such features and pre-processing steps.

%\old{We also compare our methods with baseline \Bao. Figure ~\ref{fig:offline} illustrates the variability in \Bao: the total latency is 1.67 hours after 2.2 hours of offline exploration but increases to 1.95 hours after 2.6 hours. This fluctuation suggests that \Bao may lead to query regressions, as it does not verify plans before selecting them. In terms of overall performance, \Bao takes more than 3 hours to reduce the latency to 1.5 hours, whereas \lime achieves this in 1 hour and \limeplus in just 0.5 hours. It shows that both \lime and \limeplus consistently outperform \Bao across all exploration durations\footnote{It is worth noting that the comparison is not entirely fair, as \Bao was not designed for workloads with repeated execution of the same queries, and reinforcement learning may not be ideally suited for offline exploration where regret minimization is less critical.}.}

%can make improvements quickly. But this performance improvement comes at the cost of query regressions. 
%In that sense, the gap between the \Bao line and the other techniques can be interpreted as ``the cost of zero regressions.''

\smallskip

% \begin{figure}[t]
%   \centering
%   \vspace{3ex}
%   \includegraphics[width=0.88\linewidth]{fig/greedy.pdf}
%   % \vspace{-1ex}
%   \caption{Comparison of Greedy and \lime after we add a ETL query into the Stack Workload. Note that the Default Workload Time increased from 1.46 hours to 1.62 hours. In this case, \lime consistently outperforms \greedy.}
%   \vspace{-1ex}
%   \label{fig:greedy}
% \end{figure}

\sparagraph{Does \greedy always work well?}
\label{sec:greedy}
The reader may notice that, in Figure~\ref{fig:bar}'s Stack workload results, \greedy and \lime both achieve around 1.3 hours at 1× the default workload time and approximately 1.24 hours at 2× the default workload time.
While these results suggest that \greedy and \lime have comparable performance in this instance, this does not imply that the \greedy approach is universally effective.
The \greedy method operates under the assumption that there is a correlation in the workload, such that longer-running queries have greater potential for performance improvement. However, this assumption may not hold true in real-world workloads. In fact, the performance of \greedy can significantly degrade in certain scenarios.
To illustrate this, we conduct an experiment where we add a simple ETL query to the Stack workload. This ETL query loads the joined results of the \texttt{question} and \texttt{user} tables from the Stack database into a CSV file, which takes 576.5 seconds to execute. It is obvious that changing query optimizer hints will not reduce the runtime of this ETL query. 

Figure ~\ref{fig:greedy} shows that from 0 to 3.25 hours (2x default workload time), \lime is consistently better than \greedy. This is because while \greedy persistently explore the long ETL query at each exploration step -- because it is one of the longest-running queries in the workload, \lime utilizes the predictive model to recognize that the potential gain from optimizing this query is low. As a result, \lime intelligently ignores the ETL query and explore other queries where performance improvements are more attainable. This highlights the advantage of incorporating predictive modeling into the exploration strategy.

\subsection{Overhead}
\label{sec:expr_overhead}

Figure \ref{fig:overhead} shows the cumulative overhead time cost for \lime and \limeplus during offline exploration time on the CEB workload. In this context, the offline exploration time is the time DBMS spends on executing the queries. The overhead for \lime is the computational cost of matrix completion, while for \limeplus, it includes both training the model on observed plan trees and inference on unobserved plan trees at each exploration step.
% Clearly, \limeplus requires significantly more resources than \lime. In each exploration step, \limeplus will train a model on observed plan trees and then perform inference for each unobserved plan tree, while \lime only needs to complete the matrix. 

After exploring for 6 hours, \lime incurs a total overhead of just 10 seconds, whereas \limeplus experiences an overhead of approximately 3600 seconds (60 minutes). This indicates that linear methods are at least \textbf{360x} more efficient in terms of computational resource usage. %Although two minutes of overhead may or may not seem significant, we note that \tcnn's overhead scales with the number of observed entries. 

We also experiment with \limeplus on an NVIDIA A100 GPU, tuning training and inferencing batch size to 512 for optimal performance. Even with this powerful GPU, \limeplus still requires 660 seconds (11 minutes) that add to overall query processing overhead.

In conclusion, \limeplus requires significantly more resources than \lime. Additionally, it's important to note that the implementation of \limeplus is more complex and has a large software footprint (e.g., requiring PyTorch~\cite{pytorch}) as well as feature extraction steps. On the other hand, \lime's implementation only requires near-universal linear algebra routines and straightforward filling of the observed latency time.

\subsection{Workload Shift}
\label{sec:expr_workload}

\begin{figure*}[t]
  \centering
  \begin{minipage}{0.45\textwidth}
  \includegraphics[width=1\linewidth]{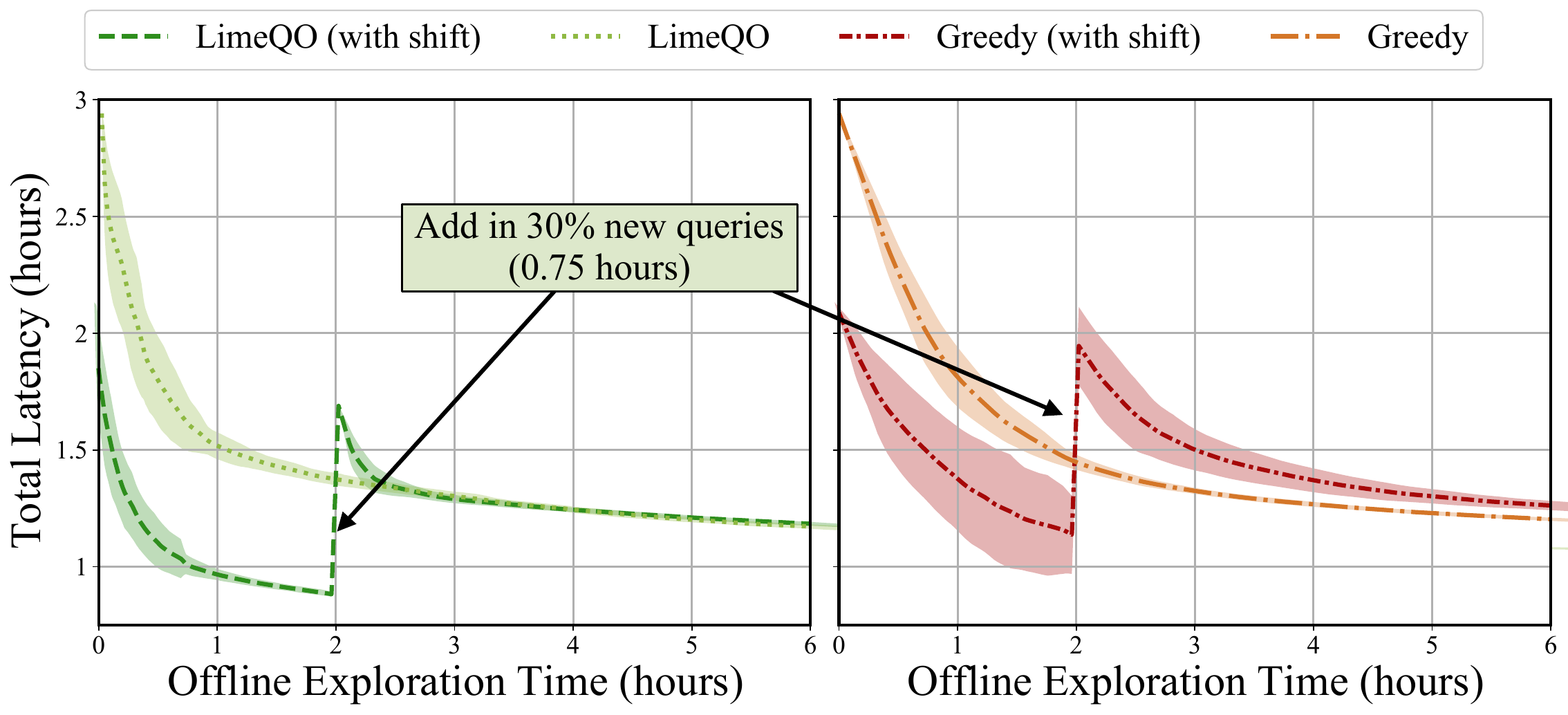}
  %\vspace{1ex}
  \caption{\lime (left figure) and \greedy (right figure) performance comparison when workload shift happens on the CEB workload. It shows \lime's ability to adapt to new queries.}
  \label{fig:newquery}
  \end{minipage}
    \hfill
    \begin{minipage}{0.245\textwidth}
        \includegraphics[width=\textwidth]{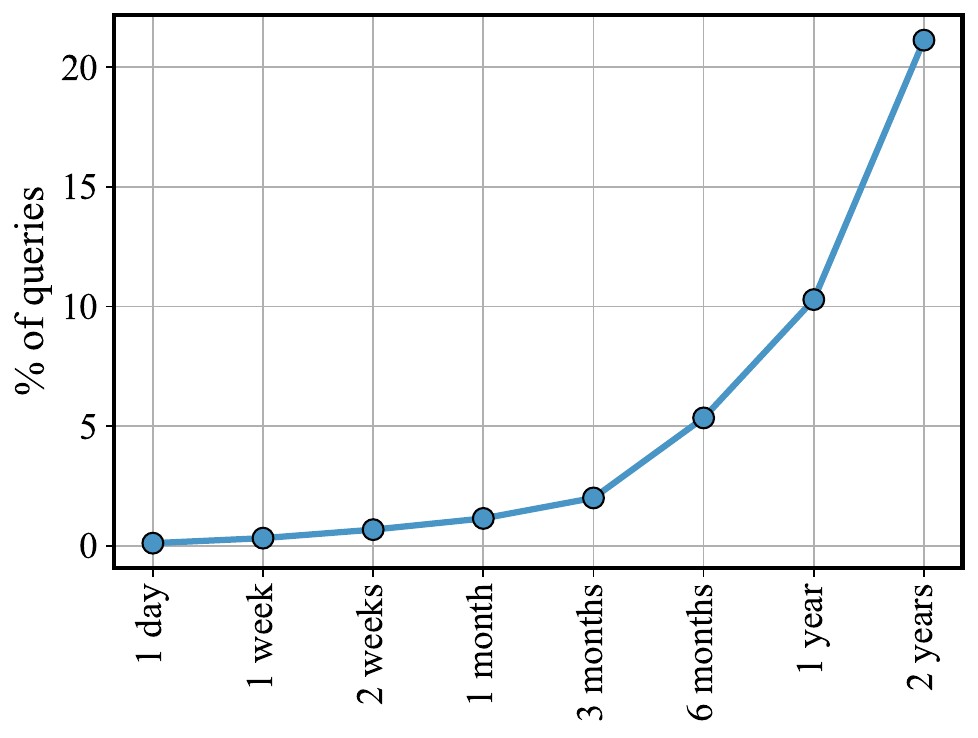}
        \vspace{-4ex}
        \caption{\new{Incremental  updates. The Y-axis is the \% of queries with a different optimal hint. Note the non-uniform X-axis.}}
        \label{fig:stack_incremental}
    \end{minipage}
    \hfill
    \hspace{0.5em}
    \begin{minipage}{0.24\textwidth}
        \includegraphics[width=\textwidth]{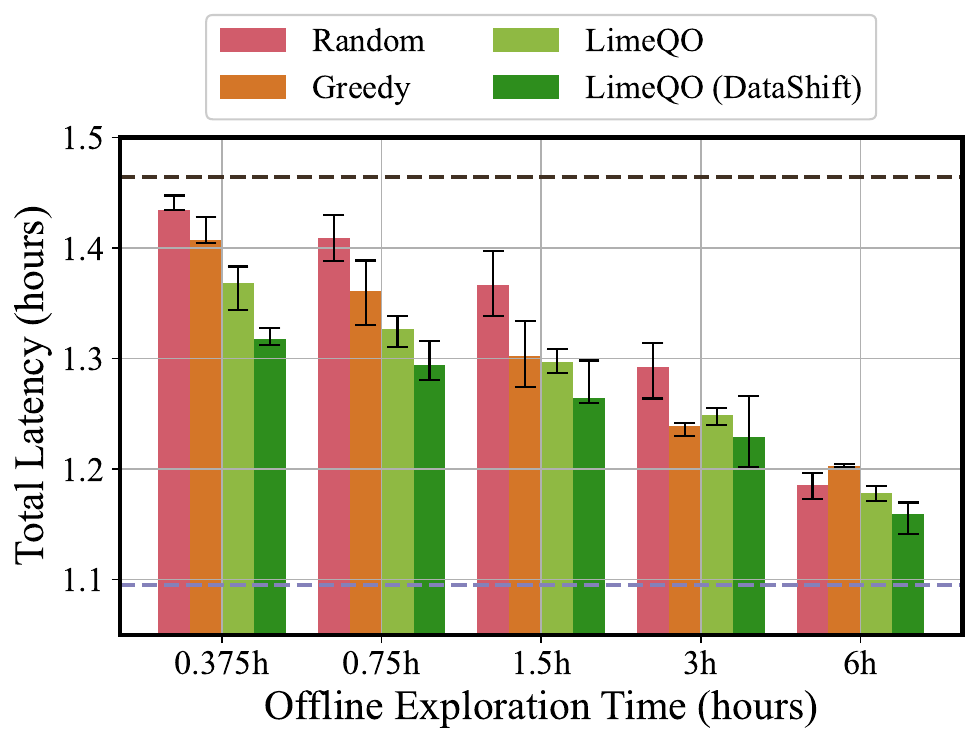}
        \caption{Data Shift on \\ Stack workload showing \\ \lime's ability  to \\ adapt to new data.}
        \label{fig:datashift}
    \end{minipage}
    % \vspace{-1ex}
\end{figure*}

\begin{figure}[t]
    \centering
    \begin{minipage}{0.23\textwidth}
        \includegraphics[width=\textwidth]{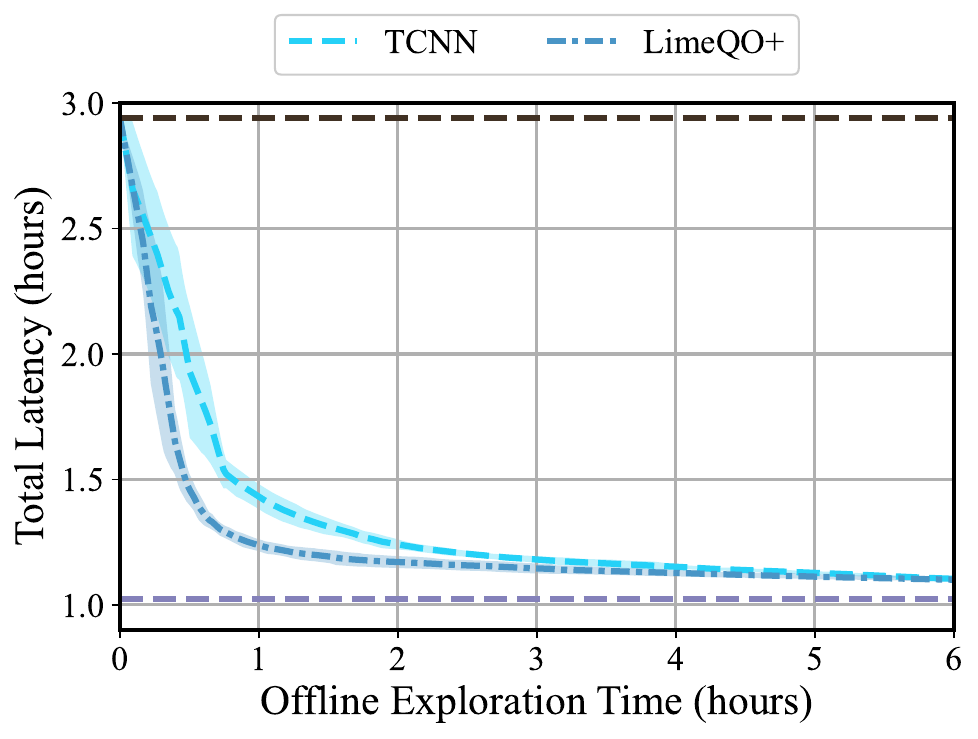}
        \caption{Total Latency \\ comparison of \tcnn VS. \\ \limeplus on CEB  \\ workload.}
        \label{fig:tcnn}
        \vspace{4ex}
    \end{minipage}
    \hfill
    \begin{minipage}{0.23\textwidth}
        \includegraphics[width=\textwidth]{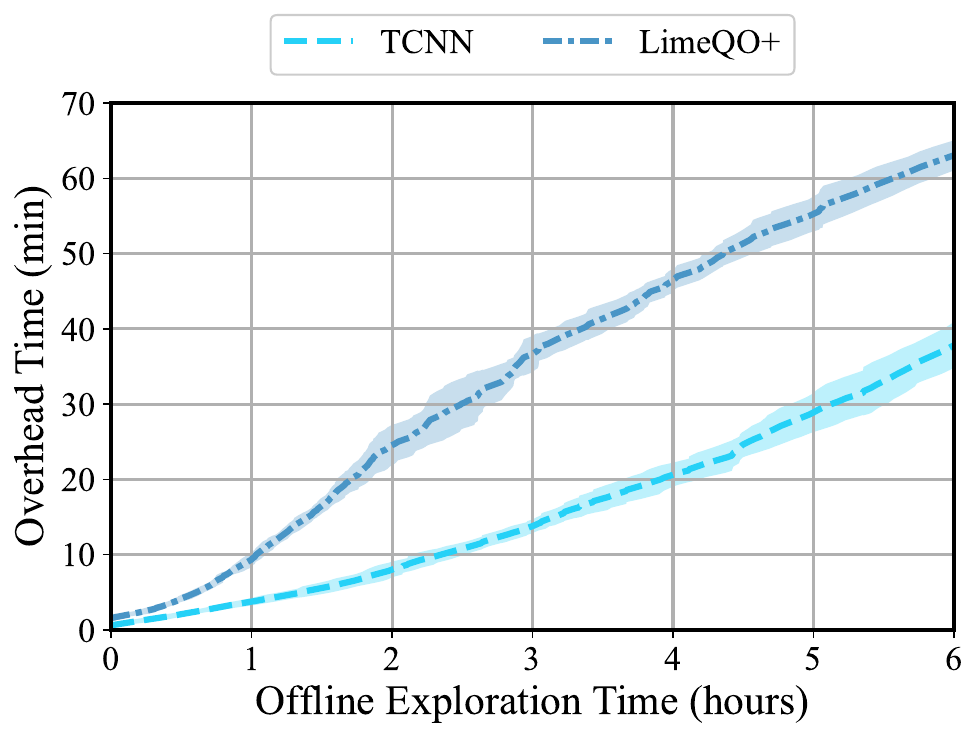}
        \caption{Overhead time \\ comparison of \tcnn VS. \\ \limeplus on CEB \\ workload.
        }
        \label{fig:tcnn-overhead}
    \vspace{4ex}
    \end{minipage}
    % \vspace{-1ex}
    \hfill
    \begin{minipage}{0.48\textwidth}
    \includegraphics[width=1\linewidth]{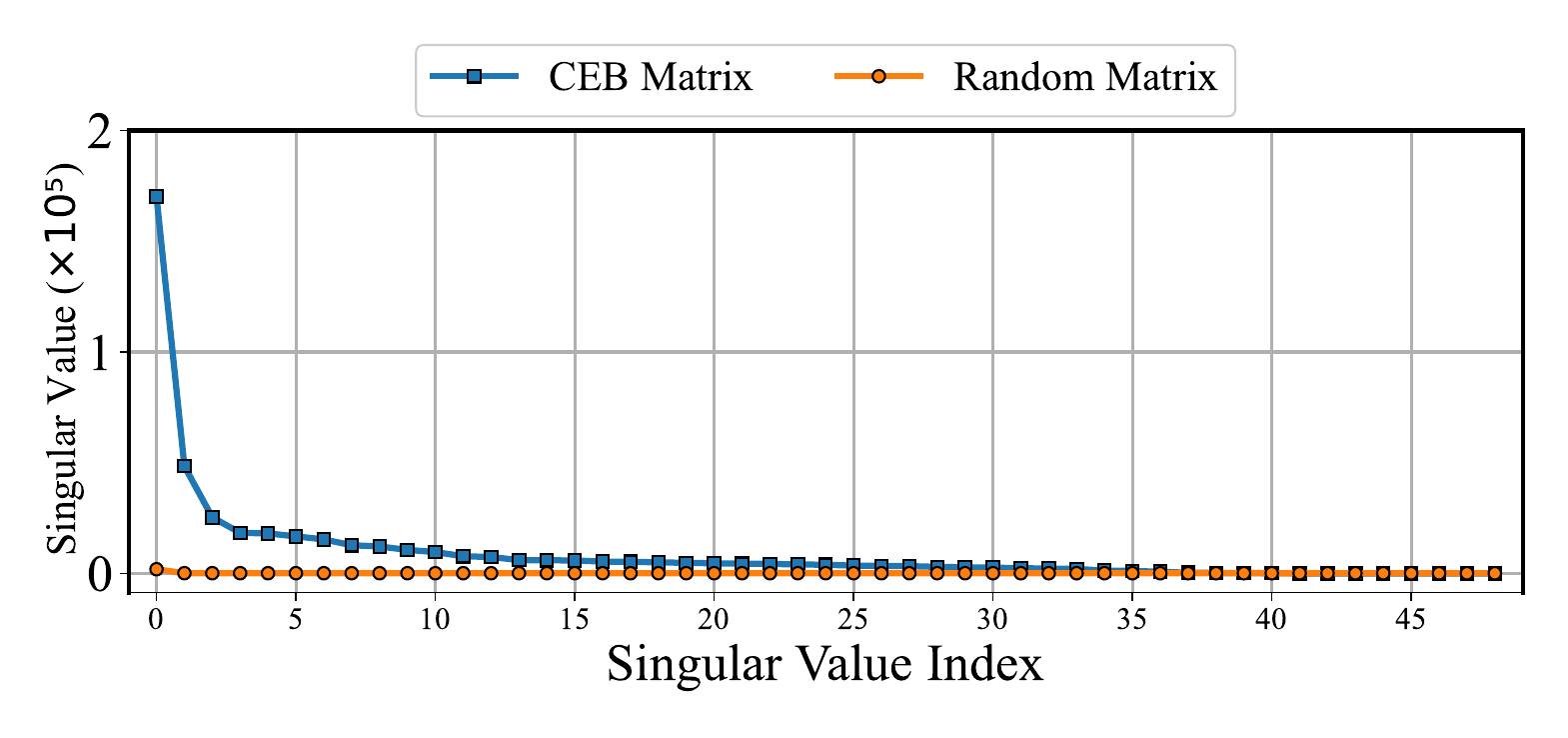}
  \vspace{-2ex}
  \caption{SVD on CEB workload matrix.}
  \vspace{-2ex}
  \label{fig:svd}
  \end{minipage}
\end{figure}

Practitioners may also concerned \lime's ability to handle new queries, particularly in the event of a workload shift. A key question is whether \lime remains robust under these conditions. 

In Figure \ref{fig:newquery}, we evaluate the performance of \lime and \greedy under a workload shift using the CEB workload. From 0 to 2 hours, we process 70\% of the full set of queries (randomly chosen). At the 2-hour mark, we introduce the remaining 30\% of queries. 

We observe that \lime reduces total latency to 0.9 hours (a 55\% gain) while \greedy only reduces to 1.2 hours (a 40\% gain). Additionally, \lime reaches the same performance level as when 100\% of the queries were available from the start after only \textbf{0.5} hours of processing the new queries. In contrast, \greedy takes longer than 4 hours to reach the performance level seen when all queries are available from the start.

This demonstrates that \lime adapts well to workload shifts while \greedy does not. It shows the advantages of \lime over simpler techniques, highlighting its robustness in dynamic environment.

\subsection{Data Shift}
\label{sec:expr_data}

% \begin{figure}[t]
%   \centering
%   \vspace{3ex}
%   \includegraphics[width=1\linewidth]{fig/datashift.pdf}
%   \vspace{-4ex}
%   \caption{Data Shift on Stack Workload showing \lime's ability to adapt to new data.}
%   \vspace{-2ex}
%   \label{fig:datashift}
% \end{figure}

Another important question we seek to answer is \lime's robustness under data drift.
To evaluate this, we utilized two versions of the Stack dataset ~\cite{bao}: one from 2017 and another from 2019. 

First, we analyze the similarities and differences between the two versions: the total default runtime increased from 1.16 hours to 1.46 hours, while the optimal runtime increased from 0.9 hours to 1.09 hours. Additionally, we examine whether the best hints for the workload have changed and find that 79\% of the workload queries maintain the same best hints. 
\new{We further evaluated incremental updates using timestamp information, with intervals ranging from 1 day to 1 week, 1 month, and 1 year. As shown in Fig ~\ref{fig:stack_incremental}, updates with 1-day intervals result in negligible changes to the optimal hints. After 1 month, 1\% of queries changed their optimal hints, 5\% after 6 months, 10\% after 1 year, and 21\% after two years.}

Applying the best hints from the 2017 dataset to the 2019 dataset reduced the total runtime from 1.46 hours to 1.26 hours, representing a 14\% reduction compared to the optimal gain of 25\%. \new{In other words, even though the hints computed for the 2017 dataset are no longer optimal in 2019, the old hints \emph{still improve latency by 14\% compared to the default optimizer}.}

Next, we simulate a complete data shift of two years after 4 hours of exploration. \new{Given that 21\% of workload queries changed the optimal hints after a 2 year data update - representing the maximum percentage observed in Fig ~\ref{fig:stack_incremental} — this experiment reflects the worst-case impact of data shift.}
Specifically, we first explore the 2017 Stack dataset for 4 hours. Then, at the 4-hour mark, we entirely shift to the 2019 dataset. We begin exploring the new dataset using the current best hints derived from the previous dataset, and then continue with the same exploration process as described in Section ~\ref{sec:linear}. Figure ~\ref{fig:datashift} shows the total workload time after spending 0.25×, 0.5×, 1×, 2×, and 4× the default workload time (1.5 hours) on the 2019 new dataset. \new{\lime is able to recover from the sudden data shift in thirty minutes, matching the performance of \lime when starting on the 2019 data.}
%It shows that \lime under this data shift consistently outperforms the version of \lime without any data shift. This demonstrates \lime's ability to adapt to new data and indicates that data drift does not significantly affect the choice of best hints.

\subsection{Ablation Study}
\label{sec:expr_ablation}
To better understand \lime and \limeplus, we analyze: (1) the role of the TCNN component in \limeplus; (2) the validity of the low-rank assumption; (3) the impact of rank selection on the performance; (4) the benefits of our censored techniques; and (5) our choice of ALS algorithm for matrix completion.

\subsubsection{\limeplus vs. \tcnn} 

% \begin{figure}[t]
% \centering
%   \vspace{3ex}
%   \includegraphics[width=1\linewidth]{fig/tcnn.pdf}
%   \vspace{-4ex}
%   \caption{\tcnn VS. \limeplus on CEB workload.}
%   \vspace{-2ex}
%     \label{fig:tcnn}
% \end{figure}

As described in Section \ref{sec:neural}, \limeplus combines linear and neural methods. To evaluate the effectiveness of this integration, we compare the performance of the pure TCNN model with \limeplus, noting that the TCNN component in both models is identical.
% To evaluate the effectiveness of this integration, we measure the performance difference between purely TCNN model and \limeplus. Notably, the TCNN component in \tcnn and \limeplus is identical.
% Figure ~\ref{fig:tcnn} shows that \limeplus consistently outperforms \tcnn throughout all the offline exploration process. After 0.75 hours, \limeplus reduces the total workload to 1.3 hours -- a 2.3x improvement -- while \tcnn only achieves a reduction to 1.55 hours, or a 1.9x gain. We also compare the overhead time  between \tcnn and \limeplus. Due to \limeplus introduces the embedding layers, \limeplus spend about 20 additional minutes of overhead after exploring 6 hours.
Figure ~\ref{fig:tcnn} shows that after 0.75 hours of optimization, \limeplus reduces latency by 56\% from 2.9 hours to 1.3 hours, while \tcnn only brings it down to 1.6 hours (a 48\% reduction). It further shows that \limeplus consistently outperforms \tcnn throughout the entire offline exploration process. This improvement can be attributed to the newly introduced features according to the low-rank property of the workload matrix, leading to more accurate predictions.  We also compare the overhead time between \tcnn and \limeplus in Figure ~\ref{fig:tcnn-overhead}. 
\limeplus spend about 20 additional minutes of overhead after exploring 6 hours. This confirms that the embedding layers in \limeplus improve performance significantly without introducing prohibitive overhead.

\subsubsection{Low-rank Structure} 
\label{sec:expr_rank}

A key assumption of \lime is that the workload matrix $\W$ has a low rank, without which matrix completion may fail to predict unobserved plans accurately~\cite{candes_power_2009}.
To validate this, we analyze the rank of the workload matrix for the CEB workload using singular value decomposition (SVD). Figure~\ref{fig:svd} presents the singular values of the complete $\W$ matrix, compared with those of a randomly generated matrix of the same shape. The workload matrix exhibits a few large singular values and many small ones, while the random matrix shows uniformly distributed singular values of similar magnitude. This observation confirms that the workload matrix can be well-approximated by a low-rank matrix, thus explaining why the ALS algorithm is effective in our scenario. For example, based on the singular values shown in Figure~\ref{fig:svd}, we find that $r < 10$ is a reasonable choice, as it captures most of the significant information while discarding smaller, less impactful singular values.
%This setting provides a good approximation of the original matrix, balancing accuracy and computational efficiency.}

% A critical assumption made by \lime is that the workload matrix $\W$ has a low rank. If $\W$ does not have low rank, it is unlikely that matrix completion will make accurate predictions for unobserved plans~\cite{candes_power_2009}. To verify this assumption, we analyze the rank of the workload matrix for the CEB workload using singular value decomposition (SVD).

% Figure~\ref{fig:svd} shows the singular values of the complete $\W$ matrix and, for comparison, a randomly generated matrix of the same shape. The singular values of the workload matrix consist of a few large values and many small values, whereas the singular values of the random matrix are uniformly distributed and of similar magnitude. This observation confirms that our workload matrix can be well approximated by a low rank matrix, thus explaining why the ALS algorithm is effective in our scenario.

\subsubsection{Rank}

\begin{figure}[t]
\centering
\begin{minipage}{0.48\textwidth}  
\includegraphics[width=1\linewidth]{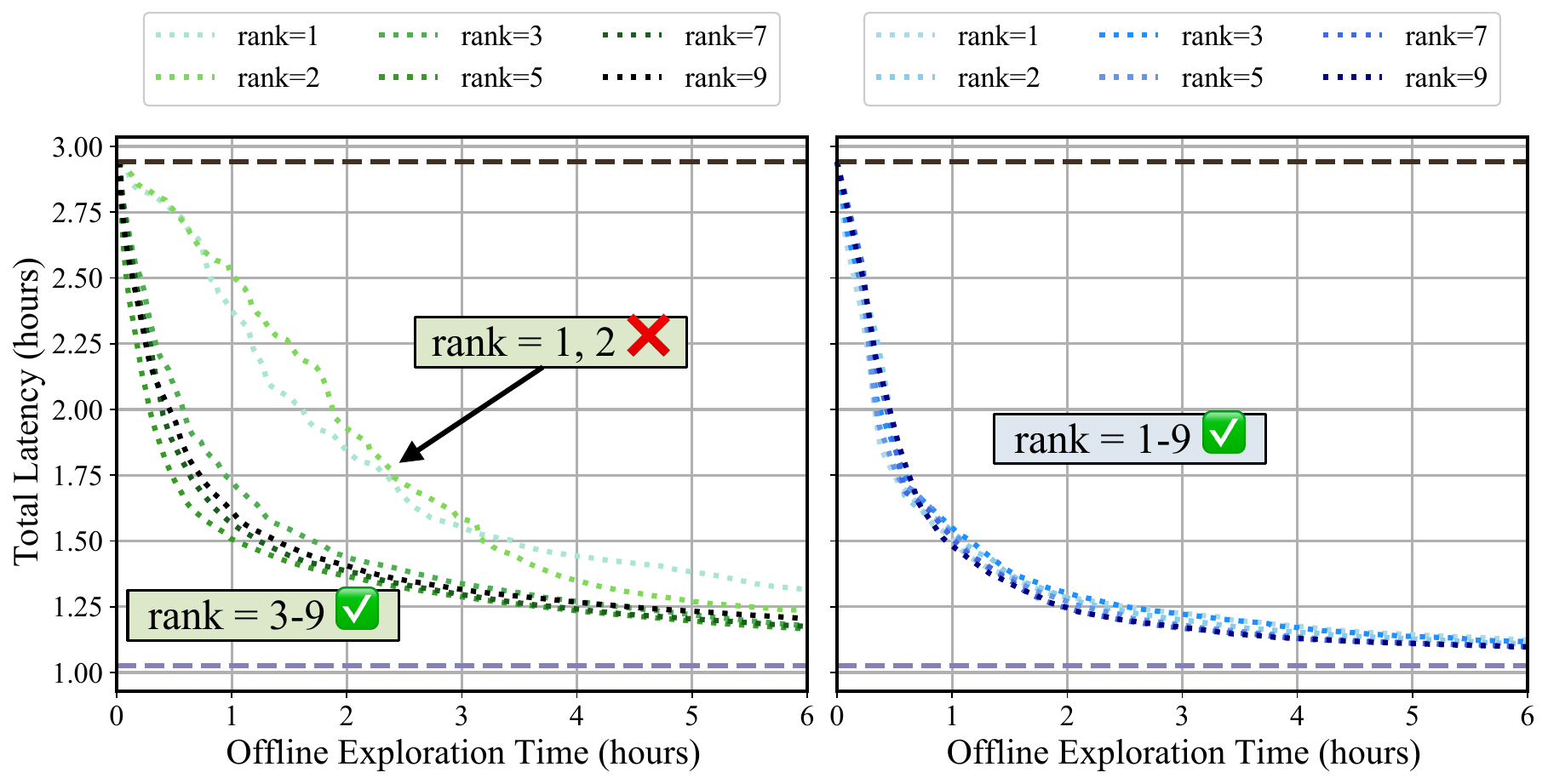}
  %\vspace{1ex}
  \caption{\lime (left figure) and \limeplus (right figure) performance on different ranks.}
    \label{fig:rank}
\end{minipage}
\hfill
\begin{minipage}{0.48\textwidth}
    \includegraphics[width=1\linewidth]{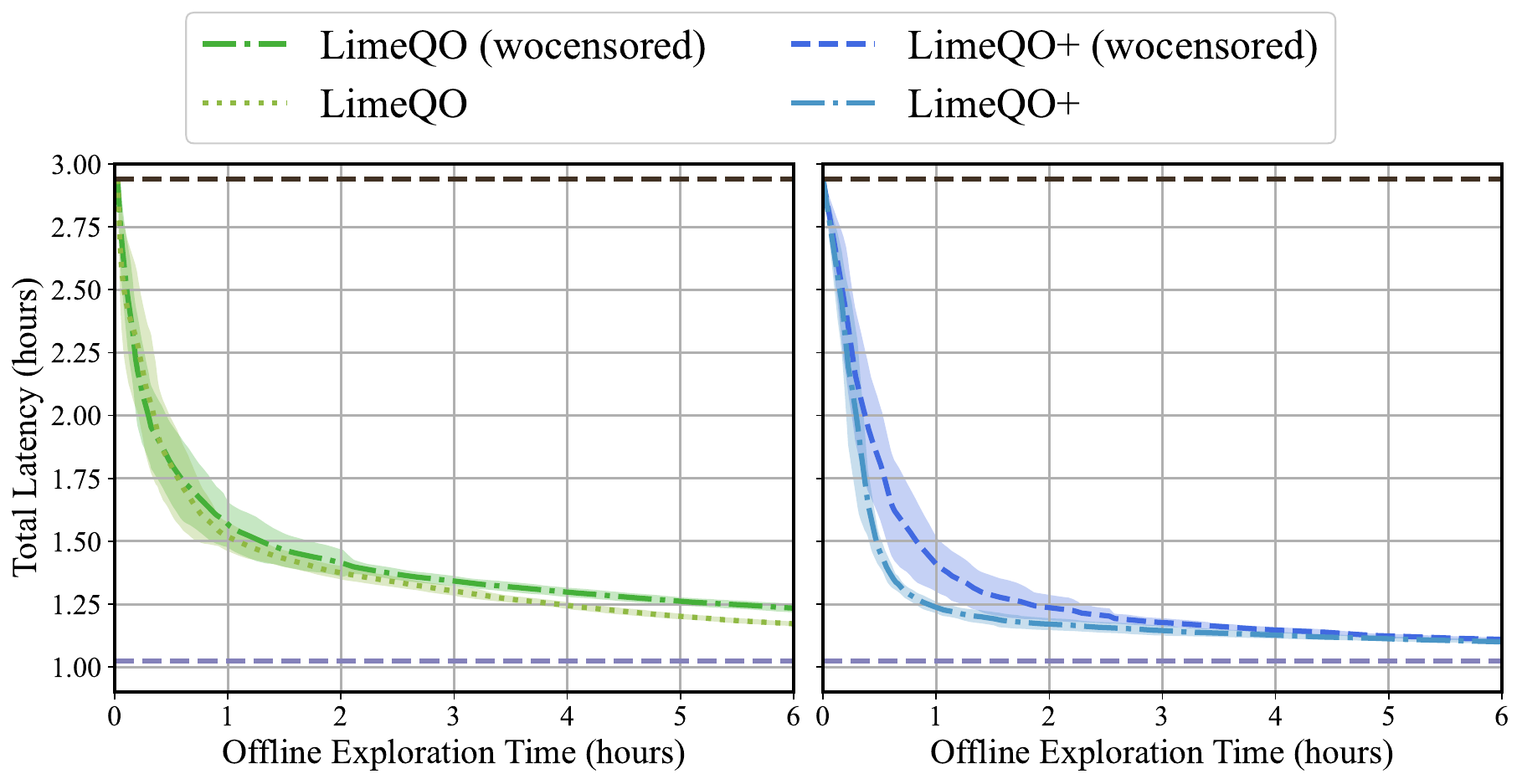}
  %\vspace{1ex}
  \caption{\lime (left figure) and \limeplus (right figure) performance with and without censored techniques.}
    \label{fig:censor}
    \end{minipage}
    \vspace{-1em}
\end{figure}

Both \lime and \limeplus require an administrator to specify the rank $r$. To evaluate how sensitive the models are to this parameter, we experimented across a range of $r$. Figure ~\ref{fig:rank} shows that \lime requires a rank greater than 2 to perform effectively, as ranks below this threshold fail to capture sufficient structure in the workload matrix. However, once the rank exceed two, the performance of \lime stabilizes and shows little variation. On the other hand, \limeplus demonstrates greater stability across different rank values, as it incorporates additional features from \tcnn, making it less sensitive to changes in $r$. This observation aligns with the findings in Section \ref{sec:expr_rank} on singular values. Ultimately, we set $r = 5$, which offers a good balance between accurately approximating the workload matrix and maintaining computational efficiency.

% Both \lime and \limeplus introduces the rank $r$, which needs to be specified in the predictive model. To assess the sensitivity of the models to this parameter, we experiment with different values of $r$ and analyze the  performance.

% Figure ~\ref{fig:rank} illustrates the impact of varying $r$ on both \lime and \limeplus. The results show that their performance remains largely invariant to the rank parameter across a broad range of values except that \lime's rank need to be larger than 2. This suggests that the models are robust to the choice of $r$ and that selecting a higher rank does not necessarily lead to significant improvements.

\subsubsection{Censored Techniques}

We introduce censored techniques in both \lime and \limeplus to handle the time-out observations. Here, we evaluate the impact of these techniques on performance. In \lime, removing the censored technique is taking out lines ~\ref{alg:censor-a} and ~\ref{alg:censor-b} in Algorithm ~\ref{alg:als}, thereby ignoring the timeout matrix. In \limeplus, it entails training the model solely on non-censored data and relying the standard MSE loss function, which does not account for timeouts.
Figure ~\ref{fig:censor} shows that applying censored techniques to \lime results in less variance and improved performance after 2 hours of exploration.  Similarly, \limeplus exhibits decreased variance and better performance.
Specifically, \limeplus with censored technique reduces the total 3-hour  workload to 1.5 hours after only 0.5 hours of exploration, whereas the version without censored technique took 0.9 hours to achieve the same reduction - a 1.8x longer exploration time.
%Notably, after 0.75 hours, \limeplus with censored techniques reduces the total workload time to 1.35 hours, whereas without these techniques, it only reduces to 1.7 hours (54\% v.s. 42\% reduction). 
These results highlight the effectiveness of censored techniques in both methods, enabling them to achieve faster convergence and more consistent performance.

% We introduced censored techniques in both \lime and \limeplus to handle the time-out observations. Here, we evaluate the impact of these techniques on performance. In \lime, removing the censored technique is taking out Lines ~\ref{alg:censor-a} and ~\ref{alg:censor-b} in Algorithm ~\ref{alg:als}, thereby ignoring the timeout matrix. In \limeplus, removing the censored technique means training the model only on non-censored data using the standard MSE loss function, thereby disregarding the censored data.

% Figure ~\ref{fig:censor} shows that applying censored techniques to \lime results in less variance and a performance improvement after 2 hours of exploring with censored techniques. Similarly, \limeplus benefits from decreased variance and demonstrates a stronger performance gain throughout the offline exploration period. Specifically, at the two-hour mark, \limeplus with censored techniques reduces the total workload time to 1.35 hours, while \limeplus without censored techniques only achieves a reduction to 1.7 hours. This illustrates that the censored techniques help both methods maintain consistent results and achieve greater performance gains.

\subsubsection{Comparisons of Matrix Completion Techniques}
\label{sec:expr_als}

% \begin{figure}[t]
%   \centering
%   \vspace{3ex}
%   \includegraphics[width=0.95\linewidth]{fig/mc.pdf}
%   \caption{Comparison of Different Matrix Completion Techniques on the JOB workload matrix. The label next to each dot indicates the proportion $p$ of the matrix that is filled.}
%   \vspace{-2ex}
%   \label{fig:mc}
% \end{figure}

\begin{figure}[t]
  \centering
  \begin{minipage}{0.65\linewidth}  % 图像的宽度，可以根据需要调整
    \centering
    \includegraphics[width=\linewidth]{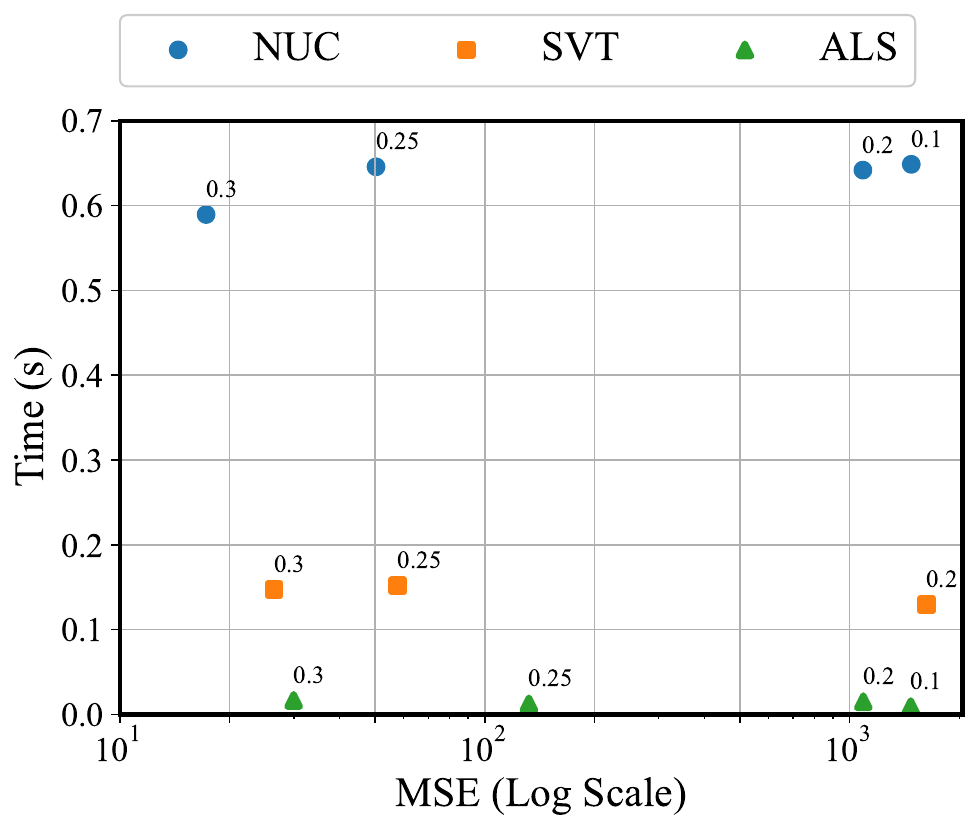}
\end{minipage}%
  \hspace{1em}  % 控制图和文字之间的间距
  \begin{minipage}{0.3\linewidth}  % caption 的宽度
    \caption{Comparison of Matrix Completion Techniques on the JOB workload matrix. The label next to each dot indicates the filled proportion $p$ of the matrix.}
    \label{fig:mc}
  \end{minipage}
\end{figure}

In this section, we explore three different matrix completion techniques and compare their accuracy and time overhead: 
\medskip

\begin{itemize}[topsep=0pt,leftmargin = 9pt]
    \item \textbf{Nuclear Norm Minimization (NUC)}  ~\cite{nuclear}: NUC recovers a low-rank matrix by minimizing the nuclear norm of the matrix subject to the constraints of observed entries, effectively leveraging the low rank property. While this approach can produce highly accurate results, it often requires substantial computational resources, especially for large datasets.
    \smallskip
    
    \item \textbf{Singular Value Thresholding (SVT)} ~\cite{svt}: SVT uses singular value decomposition (SVD) and applies a threshold to the singular values to enforce low-rank approximation. However, it may struggle with noisy data or sparse observations.
    \smallskip
    \item \textbf{Alternating Least Squares (ALS)}~\cite{als}: ALS iteratively optimizes matrix factors, makes it highly scalability. ALS is particularly effective for large-scale problems, as it can handle various forms of missing data and is less sensitive to initialization compared to other methods.
\end{itemize}
\medskip
\noindent
Figure ~\ref{fig:mc} shows that while NUC provides good accuracy, it incurs a significant computational cost, taking over 0.5 seconds on the small JOB workload matrix ($131 \times 49$). This overhead increases further for larger matrices. SVT, on the other hand, fails to handle sparse matrices effectively; its performance at $p=0.1$ is absent because it could not solve the matrix under such sparsity. In contrast, ALS balances well between accuracy and efficiency, yielding satisfactory results with the least overhead across various levels of sparsity\footnote{We do not show $p > 0.3$ because we never observed a higher $p$ in our experiments.}. Therefore, we choose ALS as our matrix completion technique in Section ~\ref{sec:linear}.

% We didn't experiment with $p > 0.3$ because, in our study, the proportion we filled in the matrices was less than 0.3 when we stopped exploring at 4 $\times$ Default Workload Time.

\begin{figure}[t]
  \begin{minipage}{0.65\linewidth}  % 图像的宽度，可以根据需要调整
    \centering
    \includegraphics[width=\linewidth]{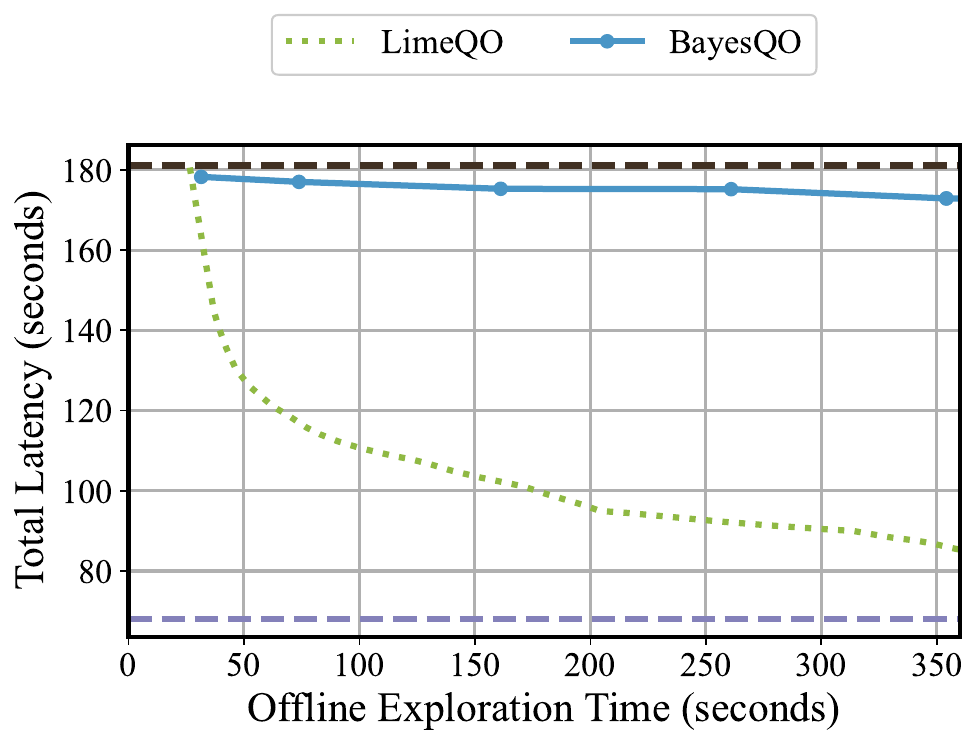}
  \end{minipage}%
  \hspace{1em}  % 控制图和文字之间的间距
  \begin{minipage}{0.3\linewidth}  % caption 的宽度
  \hspace{0.5em}  % 控制图和文字之间的间距
    \caption{\new{Comparison with BayesQO on the JOB workload. \lime significantly optimizes the workload, while BayesQO shows minimal progress.}}
    \label{fig:bayes}
    \end{minipage}
    \vspace{-1em}
\end{figure}

\new{\subsection{Comparison with BayesQO 
\label{sec:expr_bayesqo}}
%``Compare LimeQO with a workload of 100 queries and 100 minutes of optimization time to BayesQO over each of the 100 queries with 1 minute of optimization time each. We expect this to show that LimeQO can make significant progress optimizing the workload, whereas BayesQO will barely be able to get started on any particular query.''
Our work shares similarities with BayesQO~\cite{bayesqo}, a concurrent approach targeting offline query optimization. However, while BayesQO optimizes queries individually, our framework is designed to optimize an entire query workload simultaneously. To compare the two approaches, we conducted additional experiments using the JOB workload. Our method followed the approach described in Section~\ref{sec:linear}. For BayesQO, each query in the workload was allocated a fixed optimization time of three seconds, after which the total latency was calculated. As shown in Figure~\ref{fig:bayes}, our approach achieves significant progress in optimizing the workload, whereas BayesQO barely makes progress on any single query. When optimizing an entire workload, it is advantageous to allocate exploration time dynamically to the ``right'' query, as opposed to allocating exploration time evenly among all queries.
}

\section{Conclusions and Future Work}
\label{sec:conclusion}
\medskip

The question of how to effectively achieve the benefits of learned query optimization, \emph{without} suffering performance regressions or running extensive model training, has been of strong interest to the database community. Building upon the idea of \emph{offline} optimization, and using a method based on specifying \emph{optimizer hints} to modify query optimizer behavior from an external system -- this paper develops \lime: a framework for zero-regression, offline learned query optimization, without requiring extensive training, knowing specific plan features, or making assumptions about the underlying DBMS.  
Our methods are inspired by collaborative filtering, and are simple and low-overhead.  Nonetheless, our experiments validate that, with appropriate active learning strategies, we can achieve nearly as much benefit as complex deep learning approaches. %  For future work, we hope to study two futher extensions. 
We also introduced \limeplus, a more computationally expensive variant that integrates neural network techniques for faster convergence, but at the cost of higher overhead. Overall, \lime provides a practical solution for learned query optimization, ensuring efficient offline exploration without regressions and offering flexibility across different query optimization environments.

%\textcolor{orange}{%TBD: This work has presented a framework for zero-regression offline learned query optimization, and shown how simple, low-overhead linear methods can be nearly as effective as complex deep learning approaches, without requiring any plan features or making assumptions about the underlying DBMS. Several open questions and challenges remain to be investigated. 

%As future work, we hope to explore \emph{transductive} learning techniques. In ML terms, techniques such as TCNNs are called \emph{inductive} because they learn a model from training inputs and labels and then predict labels for unseen test inputs. \lime, on the other hand, is an \emph{transductive} technique, since training inputs, training labels, and test data are known in advance (only test labels are unknown). A key question is, what might a transductive TCNN look like? We would also like to explore combining MC and TCNN to create a smaller TCNN.% Or perhaps there is a middle ground between dead simple techniques like MC and more complex techniques like TCNN?

\smallskip
\textbf{Future Work.} Our existing framework relies on steering an optimizer through coarse-grained hints. In the future, we plan to investigate whether optimizers with finer-grained hints can benefit --- for instance, the optimizer for the open-source Apache Presto~\cite{sethi2019presto} distributed SQL engine. Taking this one step further, we will explore whether modern optimizer frameworks such as Apache Calcite~\cite{begoli2018apache}, used in many systems, could be extended to incorporate variations of our offline exploration model. 
We also plan to investigate techniques for \emph{online} exploration over the space of hints and plans leveraging the low-rank structure, complementing the offline exploration of our current approach. 

\new {Additionally, since the query optimizer is PostgreSQL is less sophisticated than those found in commercial systems, we plan to extend our technique to systems like SQL Server and Oracle. On one hand, these commercial systems have more powerful baseline optimizers, so improvements may be harder to find. On the other hand, these commercial systems also provide a wider variety of hints, potentially creating more opportunities for optimization.} 

\new{Evaluating our technique in the presense of data drift is also an important future direction. In this work, we looked at the impact of large data changes, but future work could evaluate performance on continuous data updates.}

\new{Finally, while we believe our incorporation of censored observations into ALS and TCNNs has been validated experimentally, future work could conduct a formal or theoretical analysis of censored techniques.}
%exploring data pipelines that have other structures like DAG?

%As future work, we plan to explore closer integration of our methods into

%Future work: finer-grained hints, extending to other systems, exploring data pipelines that have other structures, online mode

%\sparagraph{Online optimization} Although \lime supports novel queries major weakness of our preliminary version of LimeQO is handling novel queries. Adding a new, empty row to the workload matrix with no entries results in any arbitrary prediction satisfying Equation~\ref{eq:als_obj} (an under-constrained linear system). It may be possible to match novel queries to similar previously observed queries in a predictable way. But an even simpler approach could be to execute novel queries using the default optimizer first, then adding a populated row to the matrix afterward. We leave the investigation and evaluation of such techniques to future work. 

% \cgreen{future work - extend to finer grained hints, integrate with presto qo}

\section*{acknowledgment}
We thank the anonymous reviewers and our shepherd for their valuable feedback and suggestions. 

% \clearpage
\bibliographystyle{ACM-Reference-Format}
\bibliography{limeqo}


\end{document}